\documentclass[journal]{IEEEtran}

\author{ Tarun Kalluri$^{\ddagger}$, Mansi Peer$^\dagger$, Vivek Ashok Bohara$^\dagger$, Daniel B. da Costa$^\ast$, and Ugo S. Dias$^\star$\\
$^\ddagger$Department of Electronics and Electrical Engineering, IIT Guwahati, India \\
$^\dagger$Department of Electronics and Communication Engineering, IIIT Delhi, India \\
$^\ast$Department of Computer Engineering, Federal University of Cear\'{a} (UFC), Sobral, CE, Brazil \\
$^\star$Department of Electrical Engineering, University of Bras\'ilia (UnB), Brazil\\
Emails: mansip@iiitd.ac.in, tarun.05.kalluri@gmail.com,  vivek.b@iiitd.ac.in, \{danielbcosta, ugodias\}@ieee.org}
\title{Cooperative Spectrum Sharing Based Relaying Protocols With Wireless Energy Harvesting Cognitive User}
\usepackage{textcomp,graphicx,subcaption}
\usepackage[font=small,labelfont=it,justification=centering]{caption}
\usepackage{cool}
\usepackage{cite} 
\usepackage{algorithm,algpseudocode}
\usepackage{amsmath}
\usepackage{amssymb,amsthm}
\usepackage{tabularx}
\usepackage[labelfont={}]{caption}
\usepackage{epstopdf}
\usepackage{color}
\DeclareSymbolFont{Letters} {U}{zeur}{m}{n}% Euler
\DeclareMathSymbol\Gamma    {\mathalpha}{Letters}{"00}
\DeclareMathSymbol{\alpha}  {\mathalpha}{Letters}{"0B}
\DeclareMathSymbol{\beta}   {\mathalpha}{Letters}{"0C}
\DeclareMathSymbol{\gamma}  {\mathalpha}{Letters}{"0D}
\DeclareMathSymbol{\eta}    {\mathalpha}{Letters}{"11}
\DeclareMathSymbol{\theta}  {\mathalpha}{Letters}{"12}
\DeclareMathSymbol{\mu}     {\mathalpha}{Letters}{"16}
\DeclareMathSymbol{\psi}    {\mathalpha}{Letters}{"20}

\newtheorem{Prop}{Proposition}

\newcommand{\SNR}{\mathrm{SNR}}
\newcommand{\PR}{\mathrm{PR}}
\newcommand{\SR}{\mathrm{SR}}
\newcommand{\ST}{\mathrm{ST}}
\newcommand{\PT}{\mathrm{PT}}
\newcommand{\R}{\mathrm{R}}
\renewcommand{\P}{\mathrm{P}}
\newcommand{\T}{\mathrm{T}}
\newcommand{\p}{\mathrm{p}}
\newcommand{\s}{\mathrm{s}}
\newcommand{\h}{\mathrm{h}}
\renewcommand{\aa}{\mathrm{a}}
\newcommand{\bb}{\mathrm{b}}
\newcommand{\yy}{\mathrm{y}}
\newcommand{\aOne}{\mathrm{TS}-\mathrm{DF}}
\newcommand{\aTwo}{\mathrm{TS}-\mathrm{AF}}
\newcommand{\aThree}{\mathrm{PS}-\mathrm{DF}}
\newcommand{\aFour}{\mathrm{PS}-\mathrm{AF}}
\renewcommand{\Pr}{\mathrm{Pr}}

\let\oldsqrt\sqrt % it defines the new \sqrt in terms of the old one 
\def\sqrt{\mathpalette\DHLhksqrt} \def\DHLhksqrt#1#2{ %Used for new definition of square root.
\setbox0=\hbox{$#1\oldsqrt{#2\,}$}\dimen0=\ht0 \advance\dimen0-0.2\ht0 \setbox2=\hbox{\vrule height\ht0 depth -\dimen0} {\box0\lower0.4pt\box2}}

\begin{document}
  \maketitle
  
      \begin{abstract}%
      		The theory of Simultaneous Wireless Information and Power Transfer (SWIPT) in energy-constrained wireless sensor networks has attracted considerable attention from the research community due to its promising features in increasing the lifetime of devices in addition to mitigating the environment hazards caused by using conventional cell batteries. On the other hand, the advancements in the areas of cooperative spectrum sharing protocols have enabled efficient use of spectrum band between primary and secondary users. Owing to this fact, in this paper, we consider an energy-constrained secondary user which harvests energy from the primary signal and forwards this latter with the guarantee of spectrum access. Two key protocols are proposed, namely time-splitting cooperative spectrum sharing (TS-CSS) and power-sharing cooperative spectrum sharing (PS-CSS), based on time splitting and power sharing at the relay, respectively. Assuming a Nakagami-$m$ fading model, exact closed-form expressions for the outage probabilities of the primary and secondary users are derived in decode-forward (DF) and amplify-forward (AF) relaying modes. From the obtained results, it is shown that the secondary user can carry its own transmission without any adverse impact on the performance of the primary user and that the PS-CSS protocol outperforms the TS-PSS protocol in terms of outage probability over a wide range of signal-to-noise ratio (SNR). Furthermore, the effect of various system parameters, such as splitting ratio, distance between nodes and harvesting efficiency, on the system outage performance on employing the proposed protocols is investigated and several insights are drawn.%
	  \end{abstract}

	  \begin{IEEEkeywords}
	  	Cooperative spectrum sharing, nakagami fading, wireless energy harvesting, wireless sensor networks.
	  \end{IEEEkeywords}

	  \section{Introduction}
	  \label{sec:1}
	  	    \IEEEPARstart{W}{ith} the proliferation in the deployment of wireless sensor networks in the fields like automated location monitoring, indoor localization and smart building initiatives, the demand for self-sustaining and long-running sensor nodes has increased \cite{akyildiz2002wireless}. Surveillance and tracking using sensor networks have become ubiquitous, where most of the sensor nodes run on coin cell batteries \cite{anastasi2009energy}. The need to eliminate the human intervention in replacement of the batteries and to make the sensor nodes more environment friendly (no need of battery disposal) has boosted the research in the field of wireless energy harvesting. Among the possible energy sources, radio-frequency (RF) energy harvesting is of particular interest as RF signals are abundant in nature through various wireless technologies and  have the capacity to carry out information and energy transfer simultaneously \cite{xiao2014wireless}. A capacity-energy function to evaluate the fundamental performance limits between the transport of power and information simultaneously through a single noisy line has been studied in \cite{varshney2008transporting}, while the authors in \cite{zhang2013mimo} have given the performance limits of multiple-input multiple-output (MIMO) broadcasting systems for simultaneous wireless information and power transfer (SWIPT). In \cite{varshney2008transporting} and \cite{zhang2013mimo}, it was assumed that the receiver is able to extract information and power from the same received signal. The authors in \cite{zhou2013wireless} have, however, addressed practical circuit limitations for this assumption and studied receiver architectures which employ dynamic power sharing (DPS) mechanisms to perform the information decoding and power transfer. 
	  	    
	  	     Cooperative relaying schemes have been widely adopted in literature to enhance the performance of a communication system in terms of diversity, coverage extension, throughput, among others \cite{nasir2013relaying,nasir2014throughput,two-source-one-relay,medepally2010voluntary,tutuncuoglu2013cooperative}. In such schemes, the relay might first decode the received information and then transmit, which is termed as decode-and-forward (DF) relaying. Alternatively, the relay might also amplify the received signal (with a power constraint) and then transmit to the destination, which is called amplify-and-forward (AF) relaying. The authors in \cite{nasir2013relaying} and \cite{nasir2014throughput} have studied the throughput characteristics of various protocols with energy harvesting relays operating on AF and DF modes, respectively, in Rayleigh fading channels. Further relaying protocols have been explored in \cite{two-source-one-relay,medepally2010voluntary,tutuncuoglu2013cooperative} for one-way relaying systems and in \cite{chen2014wireless} for two-way relaying.

	  	    On the other hand, recent spectrum measurements have shown that although most of the spectrum band is allocated under license, the spectrum usage is very low, which has resulted in emerging research towards methods which facilitate spectrum sharing between users, also known as cooperative spectrum sharing. With the aim to improve diversity, \cite{vashistha2015outage} studies a scenario where secondary user co-exists with a primary user by acting as an altruistic relay and carries its transmission along with the transmission of a licensed primary user. The secondary user was equipped with multiple antennas thus improving the coverage and performance of both primary and secondary users. The combination of cognitive spectrum sharing and opportunistic energy harvesting has been studied in \cite{lee2013opportunistic}, where the authors have derived maximum secondary network throughput under given outage probability constraints. Outage patterns in relay assisted cognitive user which harvests energy from primary user was studied in \cite{mousavifar2014wireless} and the transmission of the secondary user was limited by the peak permissible interference power with respect to the primary transmission. 
	  	    
	  	The work in \cite{7342973} focuses on the underlay paradigm with energy harvesting secondary users where the cooperation is between the secondary source and relay, with primary user being an energy source. On the other hand, \cite{7169619} considers primary user and secondary users equipped with spectrum and energy resources, respectively, and their cooperation helps them in achieving their respective performance goals. Also \cite{6847192} proposes an optimal cooperation strategy wherein primary user switches from a non-cooperative mode to cooperative mode and secondary user harvests energy from ambient radio signals. Unlike the above works, in this paper we have proposed two cooperative spectrum sharing protocols, namely TS-CSS and PS-CSS, where the secondary user gets
access to the licensed spectrum in exchange of its assistance to primary user but the secondary user is permitted spectrum access in orthogonal time, resulting in interference avoidance. The primary user can send information to its destination only in cooperation with secondary user. Our main contributions and insights in this paper are listed as follows.
	  	    \subsection{Contributions}
	  	    \begin{itemize}
	  	    	\item We propose time-splitting and power-sharing based cooperative spectrum sharing protocols in which an energy-constrained secondary (a.k.a. cognitive) user acts as a relay for the primary signal. These novel protocols address both the problems of spectrum sharing and energy harvesting, wherein the secondary user harvests energy from the primary signal, forwards the primary signal to the intended destination, and receives spectrum access for its own transmission. The amount of energy harvested will determine the time given to the primary and secondary users for accessing the shared spectrum.

	  	    	\item Exact closed-form expressions for the outage probabilities of the primary and secondary users are derived for both protocols.

	  	    	\item The AF and DF relaying schemes have been separately considered while calculating the outage expressions. Note that both these schemes are fundamentally different based on the nature of the relaying used, as explained in \cite{su2010optimum}. AF relaying nodes are relatively simpler to build and are preferred over DF if privacy and security of the primary data are a major concern. 

	  	    	\item The analytical results are verified using numerical simulations and the effect of the various system parameters on the outage performance are studied. The optimal values of the splitting parameters and the node placement, which give the minimum outage at the destinations, are evaluated. From the obtained results, it is shown that the proposed PS-CSS protocol, which is based on power sharing receiving at ST, outperforms TS-PSS protocol, which employs time-splitting.
	  	    \end{itemize}
	  	    \subsection{Organization}
	  	    The remainder of this paper is organized as follows. In Section~\ref{sec:2}, the system model is introduced along with the key assumptions. In Sections~\ref{sec:3} and \ref{sec:4}, the TS-CSS and PS-CSS protocols are explained, respectively, and the exact closed-form expressions of the outage probabilities of both protocols are derived within AF and DF relaying. In Section~\ref{sec:5}, the proposed analytical expressions are verified through Monte Carlo simulations and the effects of various system parameters on the system outage performance are investigated, based on which several insights are attained. Section~\ref{sec:6} finally concludes the paper and summarizes the main results. 
	  	
	  	    \begin{figure}
	  	 		\centering	
	  	 		\includegraphics[width = 0.3\textwidth]{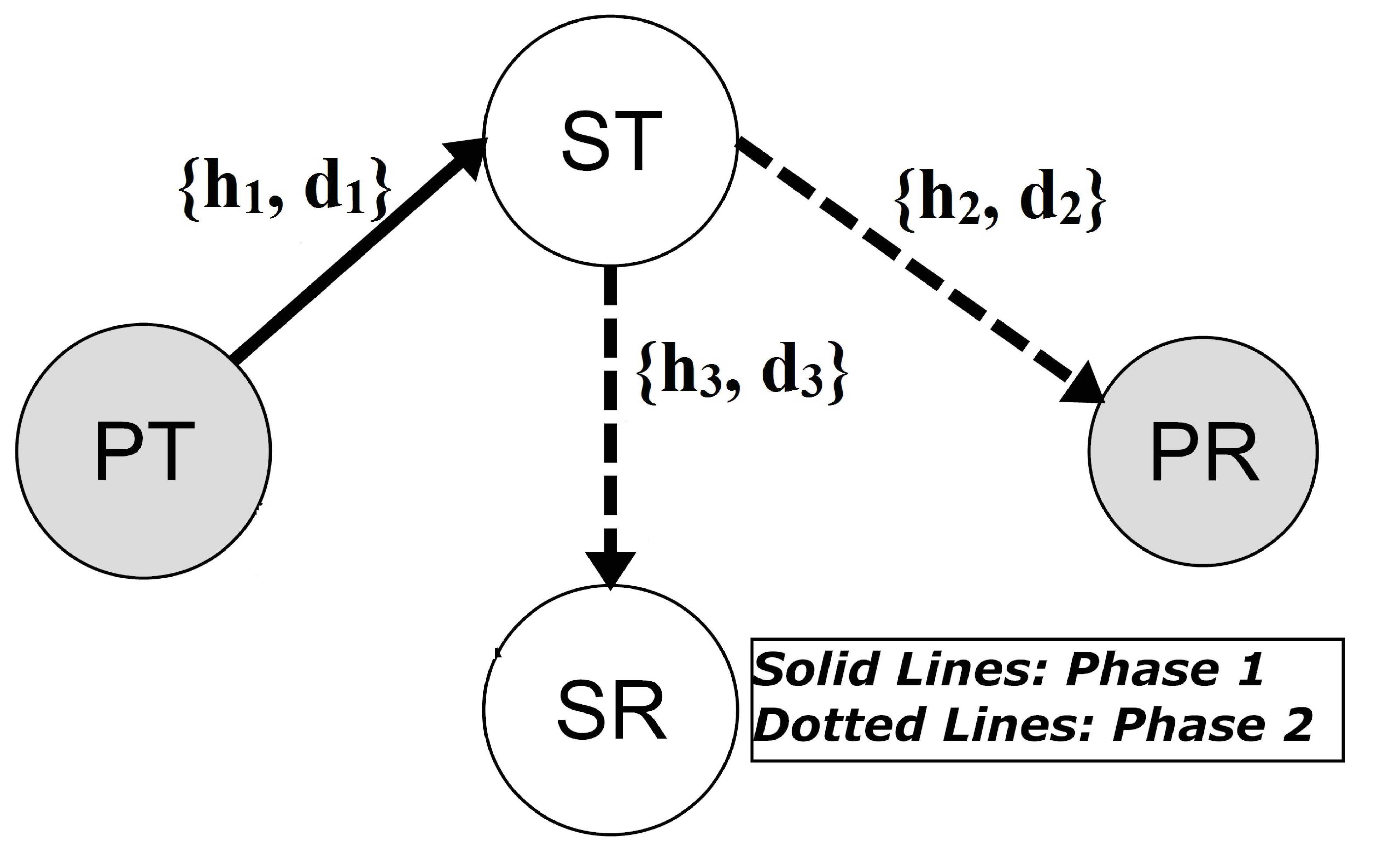}
	  	 		\caption{System Model.}
	     	\end{figure} 
	     
	  \section{System Model and Key Assumptions}
	  \label{sec:2}
			As shown in Fig. 1, the system model for our scheme contains a primary transmitter and receiver (together called primary users) and a secondary transmitter and receiver (together called secondary users or cognitive users). In Phase 1 (solid lines in Fig. 1, the primary transmitter (PT) broadcasts its signal $x_p(\mathbb{E}\left[|x_p|^2\right]=1)$ with a transmit power $\P_\p$. The secondary transmitter (ST) receives the signal, harvests energy, and processes the information. The Phase 2 (dotted lines in Fig. 1) is further subdivided into Phase 2.a and Phase 2.b. Specifically, in Phase 2.a, ST uses the harvested energy acting as an altruistic relay and forwards the primary signal to primary receiver (PR) and in Phase 2.b, ST transmits its own signal $x_s(\mathbb{E}\left[|x_s|^2\right]=1)$ towards secondary receiver (SR). The two signals sent by ST in Phase 2 are split in time, to prevent any linear inter-combination of signals, as in \cite{vashistha2015outage}, and hence avoid interference at the receivers. Thus, the performance constraint in these protocols is only due to channel fading and additive noise. 
			
			In the subsequent analysis, the performance metric is taken as the outage probabilities at the SR and PR. Specifically, we take $\R_\p$ and $\R_\s$ as the threshold rates at the primary and secondary destinations respectively, and declare an outage event if the achieved rates fall below these respective threshold rates, where we assume that both the transmitters always have some information to send. In DF relaying scheme, if ST fails to decode the primary signal, it keeps quiet in Phase 2, without transmitting its own signal. In AF relaying, however, ST performs both ST-PR and ST-SR transmission in Phase 2, irrespective of the transmission in Phase 1.

	  \subsection{Assumptions and Notation}
	  	\begin{enumerate}
	  	\item Direct link is a failure.
	  	\item No circuit level power consumption.
	  	\item No minimum power level criterion.
	  	\item All the channel links are assumed to be quasi-static block fading channels, where the channel gains are independent but not necessarily identically distributed (I.I.N.D.). In particular, we assume Nakagami-$m$ fading model\protect\footnote{$\mathit{f}_{|h_i|}(z) \sim \tfrac{2m^m}{\Gamma(m)\Omega^m_i}z^{2m-1}e^{-\frac{m z^2}{\Omega_i}}$, where $ f_{}(\cdot)$ denotes the PDF of a given random variable, $\Omega_i =$ $\mathbb{E}\left[|Z_i|^2\right]$  and $\Gamma(m) = \int_0^\infty t^{m-1} e^{-t} \mathrm{d} t$ is the Gamma function.} with channel coefficients $\h_1, \h_2, \h_3 \text{ and } \h_4$ over the channels PT-ST, ST-PR and ST-SR, respectively. Also, we represent the respective link distances by $\mathrm{d}_1, \mathrm{d}_2 \text{ and } \mathrm{d}_3$. Consequently, $|h_i| \sim \text{Nakagami}(m,\Omega_i),\quad i \in \{ 1,2,3 \}$ and \mbox{$\Omega_i=\mathrm{d}_i^{-v}$,} where $v$ is the path-loss exponent factor and $m$ is the shape parameter of Nakagami distribution. The random variables $|h_i|^2$ represented by $\gamma_i$ is Gamma distributed\protect\footnote{$\mathit{f}_{\gamma_i}(x) \sim \scriptstyle{\frac{x^{m-1} e^{-\frac{x}{\theta_i}}}{\Gamma(m) \theta^m_i}}$.} so that \mbox{$\gamma_i \sim \text{Gamma}(m,\theta_i)$}, where $\theta_i=\Omega_i/m $.
			\item We assume that there is no direct link between the nodes PT-PR so that the primary user completely relies on the secondary user for transmission, which is justified in the case of urban areas or nodes placed far apart \cite{nasir2013relaying}.  
\item We assume PU is always active \footnote{In a practical scenario, the number of PTs will not be limited to one, i.e., there will always exist some PTs in the network which are active and ST harvests energy from them \cite{7342973, 6985740}. However, in case of PT is not active, the ST remains quite. The investigation of the dynamic behavior of PU is beyond the scope of our paper but will definitely be taken into consideration in our future studies}. 
	  		\item We assume additive white Gaussian noise (AWGN) at all the receivers. The noise on channel link $i$ in transmission phase $j$ is represented as $n_{ij}$, i.e.,
	  		\begin{IEEEeqnarray}{C}
	  			n_{ij} \sim \mathcal{CN}(0,\sigma_{ij}^2) \quad i \in \{1,2,\ldots,4\} \text{ and } j \in \{1,2\}, \IEEEeqnarraynumspace
	  		\end{IEEEeqnarray}
	  		where $\sigma_{ij}^2$ denotes the AWGN variance.

	  		\item All the nodes are single-antenna devices, which is a reasonable assumption in small-sized sensor nodes \cite{akyildiz2002wireless}. We further assume harvest-store-use system of energy harvesting relaying, where the energy harvested is directly and continuously used by the relaying node (secondary transmitter in this case) \cite{sudevalayam2011energy}. The node is assumed to be equipped with an ideal storage unit with infinite battery capacity. 

	  		\item Channel state information (CSI) is assumed to be perfectly available at the receivers. For ST, we assume that the CSI is available at the receiver side but not on the transmitting side, which is in line with previous works in this area \cite{ishibashi2012energy,nasir2013wireless,chalise2012energy}.

	  	\end{enumerate}

	    \section{The TS-CSS Protocol}
	    \begin{figure}[!tp]
	  			\centering
	  			\includegraphics[width = 0.5\textwidth]{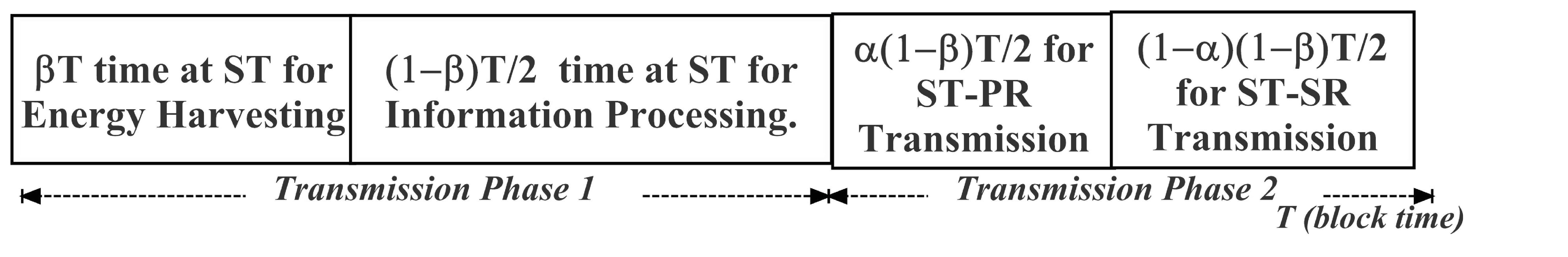}
	  			\caption{Transmission in TS-CSS protocol.}
	  			\label{fig:TS-CSS}
	  		\end{figure}

	    \label{sec:3}
	  	The TS-CSS protocol is explained in Fig. \ref{fig:TS-CSS}, where T is assumed to be the block time. In this protocol, ST harvests energy during $\beta \T$ of the time, and the remaining $(1-\beta) \T$ time is divided equally into information processing at ST and message transmission from ST to PR and SR i.e., $(1-\beta) \T/2$ time is used for information processing and $(1-\beta) \T/2$ time is used in message transmission to PR and SR from ST. Further, out of the allocated $(1-\beta) \T/2$ time for transmission,  a fraction $\alpha$ is used for ST-PR transmission and the remaining $(1-\alpha)$ for ST-SR transmission. Note that there is a two-fold time splitting, which implies to two independent parameters, $\alpha$ and $\beta$.
	  	
	  	\subsection{DF Relaying scheme}
	  	 The signal received at ST from PT in Phase 1 is given by
	  		\begin{equation}
	  			y_{\ST}=\sqrt{\P_\p}\h_1x_p + n_{11},
	  			\label{eq:yst}
	  		\end{equation}
	  	where $x_p$ is the primary signal.
	  		Since $\mathbb{E}\left[|x_p|^2\right]=1$ and $|h_1|^2 = \gamma_1$, the instantaneous SNR at ST can be written as: 
	  		\begin{equation}
	  			\SNR_\ST^{\aOne}=\frac{\P_\p \gamma_1}{\sigma_{11}^2}.
	  			\label{eq:snrst-df}	  			
	  		\end{equation}

	  		From \eqref{eq:snrst-df}, the achievable rate between PT and ST can thus be calculated as 
	  		\begin{equation*}
	  			\R_\ST^{\aOne}=\frac{(1-\beta)\T}{2}\log_2(1+\SNR_\ST^{\aOne}),
	  		\end{equation*}
	  		where the factor $(1-\beta)\T/2$ denotes the time for which ST decodes the data. Since $\beta \T$ time is used for energy harvesting, the harvested energy through this protocol is given by 
	  		\begin{equation}
	  			E_h=\zeta \P_\p \gamma_1 (\beta \T),
	  			\label{eq:eh}
	  		\end{equation}
	  		where $\zeta$ stands for the energy harvesting efficiency. Next, we denote the power allocated for ST-PR transmission by $\P_{r_1}$ and that allocated for ST-SR transmission by $\P_{r_2} $. Since the power extracted in any time $\Delta$T from energy E is given by E/$\Delta$T, it follows that
	  		\begin{IEEEeqnarray*}{rCCCl}
	  			\P_{r_1} &=& \frac{\omega E_h}{\alpha(1-\beta)\T/2} &=& \frac{2 \eta_1 \P_\p \beta}{\alpha (1-\beta)} \gamma_1\IEEEyesnumber \IEEEyessubnumber, \label{pr1-df-old} \\
	  			\P_{r_2} &=& \frac{(1-\omega) E_h}{(1-\alpha)(1-\beta)\T/2} &=& \frac{2 \eta_2 \P_\p \beta}{(1-\alpha) (1-\beta)} \gamma_1 \IEEEyessubnumber \label{pr2-df-old},
	  		\end{IEEEeqnarray*}
			  	
			where $\eta_1=\zeta\omega$, $\eta_2=\zeta (1-\omega)$  and $\omega$ is the fraction of harvested energy provided for ST-PR transmission. In the following work $\omega=1-\omega=0.5$. Hence, we can write $\eta_1$=$\eta_2$=$\eta$, and \eqref{pr1-df-old} and \eqref{pr2-df-old} will be
			  \begin{IEEEeqnarray*}{rCCCl}
	  			\P_{r_1} &=&  \frac{2 \eta \P_\p \beta}{\alpha (1-\beta)} \gamma_1\IEEEyesnumber \IEEEyessubnumber, \label{pr1-df} \\
	  			\P_{r_2} &=& \frac{2 \eta \P_\p \beta}{(1-\alpha) (1-\beta)} \gamma_1 \IEEEyessubnumber \label{pr2-df}
	  		\end{IEEEeqnarray*}
	  	In DF relaying, ST first attempts to decode the received primary signal and then transmits it towards PR in phase 2. Note that on successful decoding of  $x_p$, ST transmits primary signal during $\alpha(1-\beta)\T/2$ time and the secondary signal during $(1-\alpha)(1-\beta)\T/2$ time, and the respective received signals at PR and SR can be written, respectively, as
	  		\begin{IEEEeqnarray}{rCl}
	  			y_{\PR} &=& \sqrt{\P_{r_1}}\h_2x_p + n_{22} \nonumber, \\
	  			y_{\SR} &=& \sqrt{\P_{r_2}}\h_3x_s + n_{32} \label{eq:y2}.
	  		\end{IEEEeqnarray}
	  		On the other hand, when decoding is unsuccessful ST remains quiet during
	  $(1-\beta)\T/2$ time.	Since $\mathbb{E}\left[|x_s|^2\right]=1$, the instantaneous SNRs at PR and SR can be written as
	  		\begin{equation}
	  		\SNR_\PR^{\aOne}\,=\,\frac{\P_{r_1} \gamma_2}{\sigma_{22}^2}, \SNR_\SR^{\aOne}\,=\,\frac{\P_{r_2} \gamma_3}{\sigma_{32}^2}.
	  		\label{eq:snr1-df} 
	  		\end{equation} By substituting \eqref{pr1-df} and \eqref{pr2-df} into \eqref{eq:snr1-df}, we have that 
	  		\begin{IEEEeqnarray}{rCl}
	  			\SNR_\PR^{\aOne} &=& \frac{2\eta \P_\p \beta}{\alpha (1-\beta)\sigma_{22}^2} \gamma_1 \gamma_2 \nonumber, \\
	  			\SNR_\SR^{\aOne} &=& \frac{2\eta \P_\p \beta}{(1-\alpha) (1-\beta)\sigma_{32}^2} \gamma_1 \gamma_3. \label{eq:snr-sr1}
	  		\end{IEEEeqnarray}
	  	 	From \eqref{eq:snr-sr1}, the achievable rates at PR and SR can be calculated as
	  	 	\begin{IEEEeqnarray*}{rCl}
	  			\R_\PR^{\aOne} &=& \frac{\alpha (1-\beta)\T}{2} \log_2(1+\SNR_\PR^{\aOne}) \IEEEyesnumber \IEEEyessubnumber \label{eq:rpr},\\
	  			\R_\SR^{\aOne} &=& \frac{(1-\alpha) (1-\beta)\T}{2} \log_2(1+\SNR_\SR^{\aOne}). \IEEEyessubnumber \IEEEeqnarraynumspace \label{eq:rst}
	  		\end{IEEEeqnarray*}
	  	\begin{Prop}
	  	\label{thrm:1}
	  		The exact outage probabilities of the primary user, $\Pr_{out_1}^{\aOne}$, and that of the secondary user, $\Pr_{out_2}^{\aOne}$ in DF relaying mode for TS-CSS protocol can be derived in closed-form as
	  			\begin{IEEEeqnarray*}{rCl}
	  				\Pr_{out_1}^{\aOne} &=& 1-\left[(1-\Gamma(m,Y1))\right.\nonumber\\
	  						   && \left.\left(1-\frac{1}{\Gamma(m)^2}\,\MeijerG[\mu]{1}{1}{m,m}{0}{\frac{Z1}{\theta_1 \theta_2}}\right)\right], \IEEEeqnarraynumspace \IEEEyesnumber \IEEEyessubnumber \label{eq:pout1-df} \\
	  				\Pr_{out_2}^{\aOne} &=& 1-\left[(1-\Gamma(m,Y1))\right.\nonumber \\
	  						   && \left.\left(1-\frac{1}{\Gamma(m)^2}\,\MeijerG[\mu]{1}{1}{m,m}{0}{\frac{Z2}{\theta_1 \theta_3}}\right)\right], \IEEEeqnarraynumspace \IEEEyessubnumber \label{eq:pout2-df}
	  			\end{IEEEeqnarray*}%
	  			\textnormal{where}, \\
	  			\begin{tabularx}{0.47\textwidth}{@{}XX@{}}
 					 \begin{IEEEeqnarray*}{rCl}
  					  \mathrm{Y1} &=& \tfrac{\psi_1 \sigma_{11}^2}{\P_\p \theta_1}, \IEEEyesnumber \IEEEyessubnumber \label{y1-df}
   						\\ 
	 				  \mathrm{Z1}&=&\tfrac{\alpha(1-\beta)\sigma_{22}^2 \psi_2}{2\eta \P_\p \beta}, \IEEEyessubnumber \label{z1-df}
	 					\\
	  				  \mathrm{Z2}&=&\tfrac{(1-\alpha)(1-\beta)\sigma_{32}^2\psi_3}{2\eta\P_\p \beta}, \IEEEeqnarraynumspace \IEEEyessubnumber \label{z2-df}
  					\end{IEEEeqnarray*}
  					&
  					\begin{IEEEeqnarray*}{rCl}
 						 \psi_1 &=& 2^{{\frac{2 \R_\p}{(1-\beta)\T}}}-1, \IEEEyessubnumber \label{eq:psi1-df}
	  					 \\
	 					 \psi_2 &=& 2^{{\frac{2 \R_\p}{\alpha (1-\beta) \T}}}-1, \IEEEyessubnumber \label{eq:psi2-df}
	 					 \\
	  					 \psi_3 &=& 2^{{\frac{2 \R_\s}{(1-\alpha)(1-\beta) \T}}}-1. \IEEEeqnarraynumspace \IEEEyessubnumber \label{eq:psi3-df}
  					\end{IEEEeqnarray*}
				\end{tabularx}
			\end{Prop}

	  		\begin{IEEEproof}
	  			Please, refer to Appendix~\ref{apndx1}. \qedhere
	  		\end{IEEEproof}
	  The function $\Gamma(.,.)$ is the normalized lower incomplete Gamma function\protect\footnote{$\Gamma(m,x)=\displaystyle{\dfrac{\int_0^x{t^{m-1} e^{-t}\,\text{d}t}}{\Gamma(m)}}$.}. The \mbox{Meijer-G} function $G^{\,m,n}_{\,p,q}$, defined in \cite[eq.9.301]{jeffrey2007table}, is given below
	  		\begin{multline}
	  			 \MeijerG[a,b]{n}{p}{m}{q}{z}= \\ 
	  			 \frac{1}{2 \pi i}\int{\frac{\prod_{j=1}^{n}\Gamma(1-a_j+s)\cdot\prod_{j=1}^{m}\Gamma(b_j-s)}
						 	  			    {\prod_{j=n+1}^p\Gamma(a_j-s)\prod_{j=m+1}^q\Gamma(1-b_j+s)} z^{s}\textrm{d}s}.
				 \label{eq:meijerG}
	  		\end{multline}
	  \subsection{AF relaying scheme}
	     Note that, different from DF relaying, in AF relaying mode ST amplifies the received signal from PT in Phase 1, and then transmits the amplified signal to PR. The signal received at PR can be written as
	  		\begin{equation}
	  			y_{\PR} = \frac{\sqrt{\P_{r_1}}}{\sqrt{\P_\p \gamma_1+\sigma_{11}^2}} y_{\ST} \h_2 + n_{22},
	  			\label{eq:ypr-ts-af}
	 		\end{equation}%
where the term $\sqrt{\P_\p \gamma_1+\sigma_{11}^2}$ in the denominator means the power constraint factor at the relay\cite{medepally2010voluntary}, chosen to ensure that the transmit power averaged over noise and interference is $\P_{r_1}$. By replacing (2) into \eqref{eq:ypr-ts-af}, we get 
	 		\begin{equation}
	 			y_{\PR} = \underbrace{\frac{\sqrt{\P_{r_1} \P_\p}}{\sqrt{\P_\p \gamma_1 + \sigma_{11}^2}} \h_1 \h_2 x_p}_\text{signal component} + \underbrace{\frac{\sqrt{\P_{r_1}}}{\sqrt{\P_\p \gamma_1 + \sigma_{11}^2}} n_{11} \h_2 + n_{22}}_\text{noise component},
	 		\end{equation}
	 	where the instantaneous SNR at PR can be attained from \eqref{eq:snrpr-tsaf} after the appropriate substitutions and arrangements as
	 		\begin{IEEEeqnarray}{rcl}
	 			\SNR_\PR^{\aTwo} &=& \frac{\left(\displaystyle{\frac{2 \eta \P_\p \beta}{\alpha(1-\beta)\sigma_{22}^2}\gamma_1 \gamma_2}\right)\left(\displaystyle{\frac{\P_\p}{\sigma_{11}^2}\gamma_1}\right)}{\left(\displaystyle{\frac{2 \eta \P_\p \beta}{\alpha(1-\beta)\sigma_{22}^2}\gamma_1 \gamma_2} \right) + \left(\displaystyle{\frac{\P_\p}{\sigma_{11}^2}\gamma_1}\right)+ 1}. \IEEEeqnarraynumspace
	 			\label{eq:snrpr-tsaf}
	 		\end{IEEEeqnarray}
	 	Thus, the achievable rate at PR can be written as
	 	\begin{IEEEeqnarray}{rCl}
	 		\R_\PR^{\aTwo} &=& \frac{\alpha (1-\beta)\T}{2} \log_2(1+\SNR_\PR^{\aTwo}).
	 		\label{eq:rpr-tsaf}
	 	\end{IEEEeqnarray}
	 	Since the secondary transmission at ST towards SR is independent of the mode of relaying used, it follows the same pattern as in DF relaying. Therefore, the received signal at SR, $y_{\SR}$, is given by \eqref{eq:y2} and the instantaneous SNR at SR can be written as
	 	\begin{IEEEeqnarray}{rCl}
	 		\SNR_\SR^{\aTwo} &=& \frac{2\eta \P_\p \beta}{(1-\alpha) (1-\beta)\sigma_{32}^2} \gamma_1 \gamma_3, 
	 	\end{IEEEeqnarray}
	    yielding in the following rate at SR
	    \begin{IEEEeqnarray}{rCl}
	    	\R_\SR^{\aTwo} &=& \frac{(1-\alpha) (1-\beta)\T}{2} \log_2(1+\SNR_\SR^{\aTwo}).
	    \end{IEEEeqnarray}

	    We note that since there will always be transmission in Phase 2 of AF relaying scheme, it makes better utilization of the block time T. 
	 		\begin{Prop}
	 		\label{thm1-af}
	 			The exact outage probabilities of the primary user, $\Pr_{out_1}^{\aTwo}$, and of the secondary user, $\Pr_{out_2}^{\aTwo}$ in AF relaying mode for TS-CSS protocol can be attained as
	 			\begin{IEEEeqnarray*}{lll}
	 				\Pr_{out_1}^{\aTwo} &=&
	 				1-\int_\mathrm{Y1}^\infty \tfrac{\yy^{(m-1)} e^{-\yy}}{\Gamma(m)} \Gamma_u\left(m,\tfrac{\psi_1(\bb\theta_1 \yy + 1)}{\theta_1 \theta_2 (\aa\bb\theta_1 \yy^2 - \psi_1 \aa \yy)}\right) \mathrm{d}\yy, \\ \IEEEyesnumber \IEEEyessubnumber \label{eq:pout1-af} \IEEEeqnarraynumspace\\
	 				\Pr_{out_2}^{\aTwo} &=&
	 				\frac{1}{\Gamma(m)^2} \ \MeijerG[\mu]{1}{1}{m,m}{0}{\frac{Z2}{\theta_1 \theta_3}}, \IEEEeqnarraynumspace \IEEEyessubnumber \label{eq:pout2-af} \IEEEeqnarraynumspace
	 			\end{IEEEeqnarray*}%
	 			\textnormal{where},
	  			
	  			\begin{tabularx}{0.45\textwidth}{@{}XX@{}}
 					 \begin{IEEEeqnarray*}{lll}
  					  \aa &=& \tfrac{2 \eta \P_\p \beta}{\alpha(1-\beta)\sigma_{22}^2}, \IEEEyesnumber \IEEEyessubnumber \label{a-af}
  					  \\
  					  \mathrm{Y1} &=& \tfrac{\psi_1 \sigma_{11}^2}{\P_\p \theta_1}, \IEEEyessubnumber \label{eq:y1-af}
	  				  \\ 
	 				  \mathrm{Z2} &=& \tfrac{(1-\alpha)(1-\beta)\psi_2\sigma_{32}^2}{2 \eta \P_\p \beta}, \IEEEeqnarraynumspace\IEEEyessubnumber \label{eq:z2-af}
  					\end{IEEEeqnarray*}
  					&
  					\begin{IEEEeqnarray*}{lll}
  						\bb &=&\tfrac{\P_\p}{\sigma_{11}^2}, \IEEEyessubnumber\label{b-af}
 						 \\
	 					\psi_1 &=& 2^{{\frac{2 \R_\p}{\alpha (1-\beta) \T}}}-1, \IEEEyessubnumber \label{eq:psi1-af}
	 					\\
	 					\psi_2 &=& 2^{{\frac{2 \R_\s}{(1-\alpha) (1-\beta) \T}}}-1. \IEEEeqnarraynumspace \IEEEyessubnumber \label{eq:psi2-af}
  					\end{IEEEeqnarray*}
				\end{tabularx}
	 		\end{Prop}

	 		\begin{IEEEproof}
	  			Please, refer to Appendix~\ref{apndx2}. \qedhere
	  		\end{IEEEproof}
	  	 The function $\Gamma_u(.,.)$ is the upper normalized incomplete Gamma function\protect\footnote{$\Gamma_u(m,x) + \Gamma(m,x) = 1$.}.
	  \begin{figure}[!tp]
	  	   		\centering	  			
	  			\includegraphics[width = 0.4\textwidth]{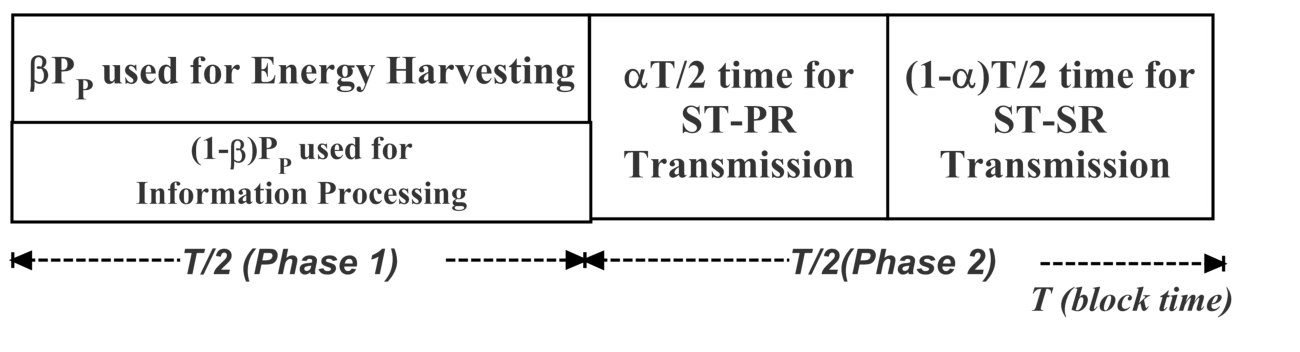}
	  			\caption{Transmission in PS-CSS protocol.}
	  			\vspace*{-0.5cm}
	  			\label{fig:PS-CSS}
	  	\end{figure}
	  \section{The PS-CSS scheme}
	  \label{sec:4}

	  		The PS-CSS protocol is explained in Fig. \ref{fig:PS-CSS}, where T is assumed to be the block time. In Phase 1, during T/2, the ST uses $\beta$ of the power from the received signal ($\beta\P_\p$) for energy harvesting and $(1-\beta) \P_\p$ for information decoding. In Phase 2, the next T/2 time is split in the ratio $\alpha:(1-\alpha)$, where ST-PR transmission is carried out in the first $\alpha \T/2$ and ST-SR transmission in the remaining $(1-\alpha)\T/2$ time. Note that again, the signals are split in time in Phase 2 to avoid any interference at the receivers. 

	  	\subsection{DF Relaying Scheme}
	  	 In DF relaying scheme for PS-CSS protocol, the signal received at ST from PT in Phase 1 is given by \eqref{eq:yst}. The ST then splits the signal between the energy harvester and information decoder. The signal received at the information receiver is given by 
	  		\begin{equation}
	  			y_{\ST}^{\,\,\prime}=\sqrt{(1-\beta)}y_{\ST}=\sqrt{(1-\beta)\P_p} h_1 x_p + n_{11}.
	  			\label{eq:yst-ps}
	  		\end{equation} 
	  	We have assumed that the noise factor is not affected by the power sharing, which is generally the case and justified in \cite{ding2014power}. The  instantaneous SNR at ST is given by: 
	  		\begin{equation}
	  			\SNR_\ST^{\aThree}=\frac{(1-\beta) \P_\p \gamma_1}{\sigma_{11}^2}.	
				\label{eq:snr-st-2} 			
	  		\end{equation}
	  	From above, the rate at ST, $\R_\ST^{\aThree}$ can be written as
	  		\begin{equation}
	  		\R_\ST^{\aThree}=\frac{\T}{2}\log_2(1+\SNR_\ST^{\aThree}).
	  		\end{equation}
	  	The energy harvested at ST is given by $E_h'=\zeta \beta \P_\p \gamma_1 (\T/2)$. Next, we denote the power allocated in PS-CSS protocol for ST-PR transmission by $\P_{r_1}^{\prime}$ and that allocated for ST-SR transmission by $\P_{r_2}^{\prime}$%\protect\footnote{The $\P_{r_1}$ or $\P_{r_2}$ refers to which mode in which protocol should be clear from the context}. %
	  	Therefore,
	  		\begin{IEEEeqnarray}{rCCCl}
	  			\P_{r_1}^{\prime} &=& \frac{\omega E_h'}{\alpha \T/2} &=& \frac{\eta \beta \P_\p}{\alpha} \gamma_1 \nonumber, \\
	  			\P_{r_2}^{\prime} &=& \frac{(1-\omega) E_h'}{(1-\alpha) \T/2} &=& \frac{\eta \beta \P_\p}{(1-\alpha)} \gamma_1, \label{eq:prdf2-ps}
	  		\end{IEEEeqnarray}
	  	where $\eta=\zeta \omega$ and $\omega=0.5$.
	  	Similar to TS-CSS protocol, ST first transmits the signal $x_p$ with power $\P_{r_1}^{\prime}$ which is received by PR and then transmits its own signal $x_s$ with power $\P_{r_2}^{\prime}$, which is received by SR. Therefore, the signals received at PR and SR in phase 2 are given, respectively, by
	  	\begin{IEEEeqnarray}{rCl}
	  		y_{\PR} &=& \sqrt{\P_{r_1}^\prime}\h_2x_p + n_{22}, \nonumber \\
	  		y_{\SR} &=& \sqrt{\P_{r_2}^\prime}\h_3x_s + n_{32}. \label{eq:ysr-ps}
	  	\end{IEEEeqnarray}
	  	yielding in the following instantaneous SNRs

	  		\begin{IEEEeqnarray}{rCl}
	  			\SNR_\PR^{\aThree} &=& \frac{\eta \P_\p \beta}{\alpha \sigma_{22}^2} \gamma_1 \gamma_2, \nonumber \\
	  			\SNR_\SR^{\aThree} &=& \frac{\eta \beta \P_\p}{(1-\alpha) \sigma_{32}^2} \gamma_1 \gamma_3.
	  			\label{snrdf2}
	  		\end{IEEEeqnarray}%
	  	 The corresponding rates at PR and SR can be given from \eqref{snrdf2} as
	  	 	\begin{IEEEeqnarray}{rCl}
	  			\R_\PR^{\aThree} &=& \frac{\alpha \T}{2} \log_2(1+\SNR_\PR^{\aThree}),\nonumber \\
	  			\R_\SR^{\aThree} &=& \frac{(1-\alpha) \T}{2} \log_2(1+\SNR_\SR^{\aThree}). \label{eq:raf}
	  		\end{IEEEeqnarray}%
	  	\begin{figure*}
	  	   \centering
	  	   \begin{subfigure}[!t]{0.5\textwidth}
	  	   		\centering
        		\includegraphics[width=\textwidth]{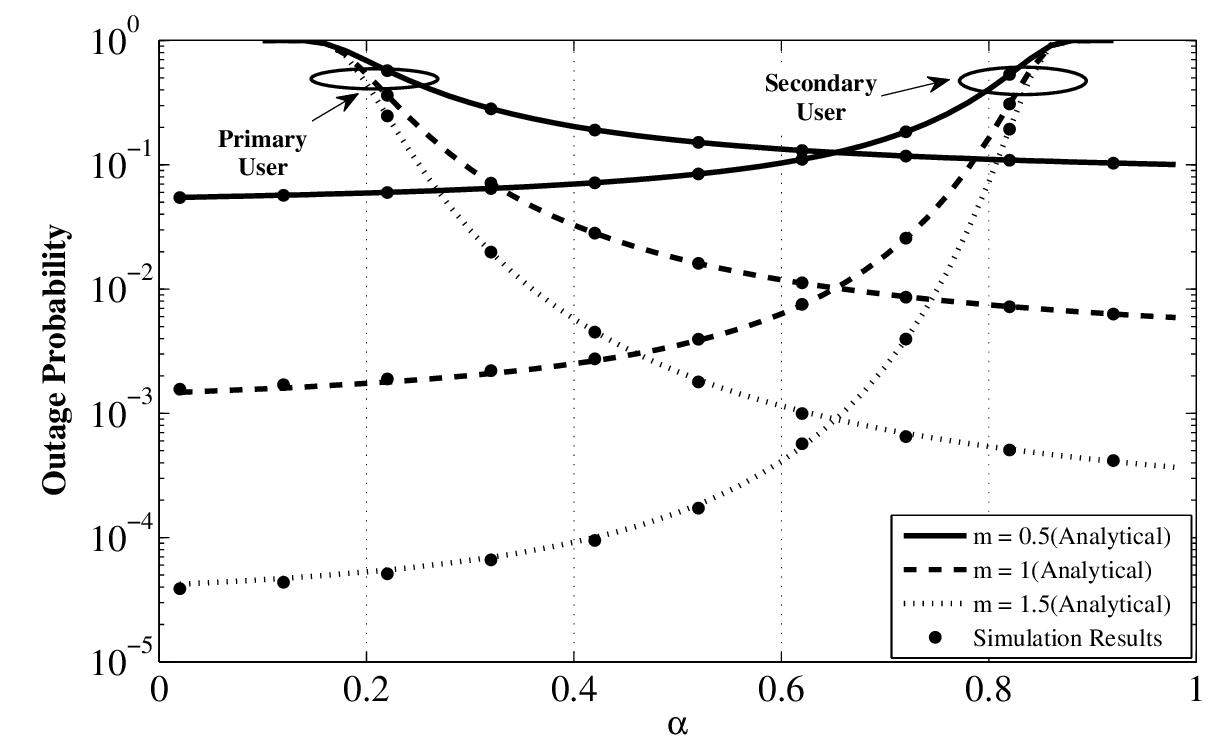}
        		\caption{}
        		\label{fig:va1-df}
	  	   \end{subfigure}%
	  	   ~%
	  	   \begin{subfigure}[!t]{0.5\textwidth}
	  	   		\centering
        		\includegraphics[width=\textwidth]{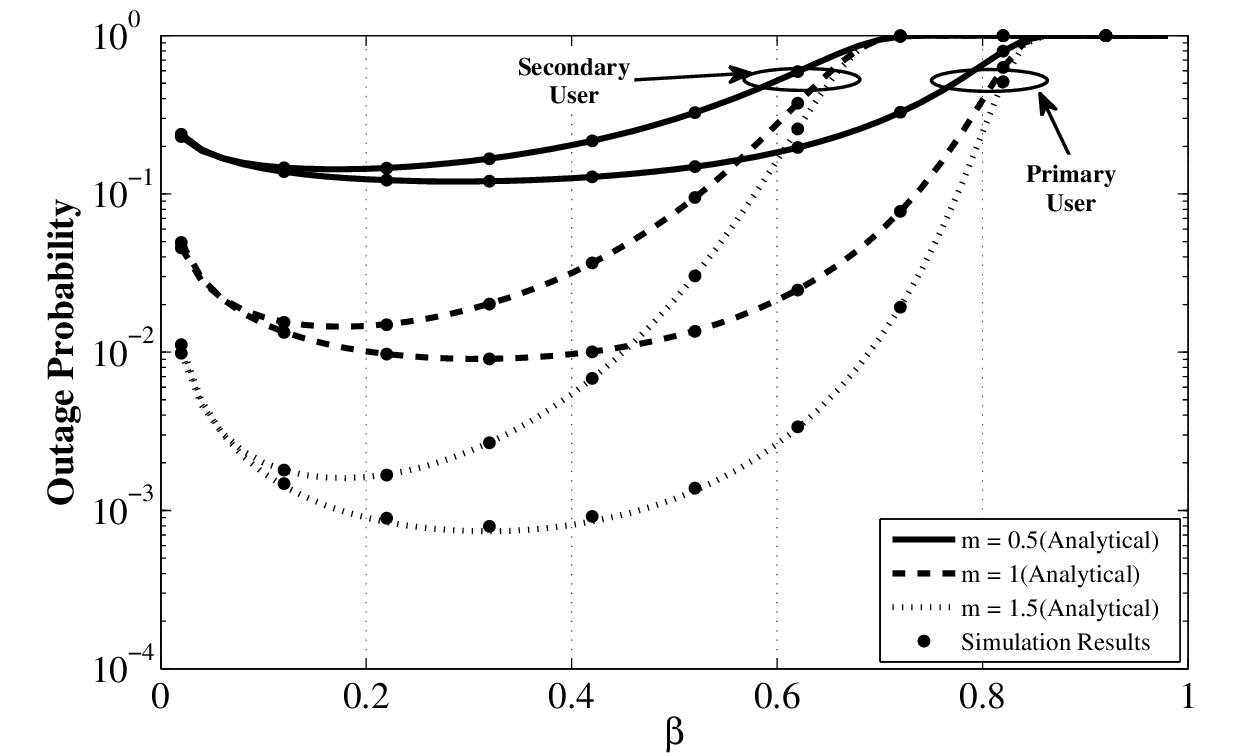}
        		\caption{}
        		\label{fig:vb1-df}
        	\end{subfigure}
	  	
	  	   \begin{subfigure}[!t]{0.5\textwidth}
	  	   		\centering
        		\includegraphics[width=\textwidth]{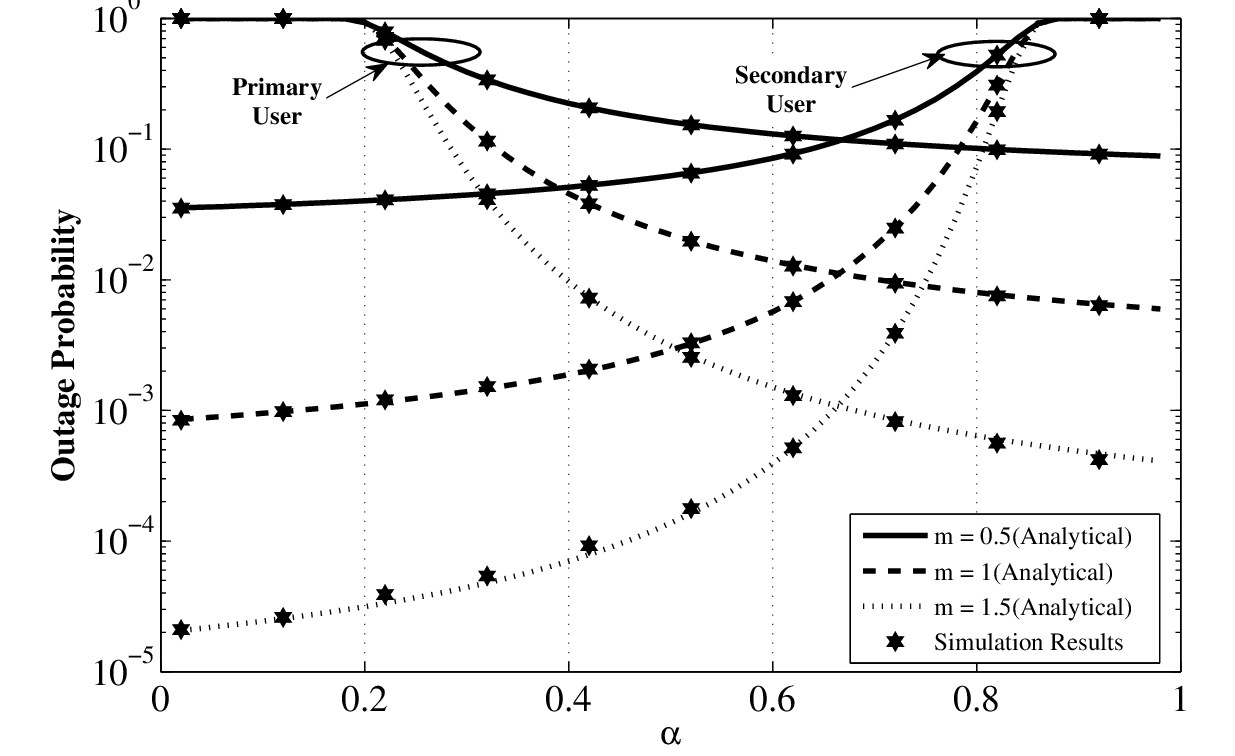}
        		\caption{}
        		\label{fig:va1-af}
	  	   \end{subfigure}%
	  	   ~%
	  	   \begin{subfigure}[!t]{0.5\textwidth}
	  	   		\centering
        		\includegraphics[width=\textwidth]{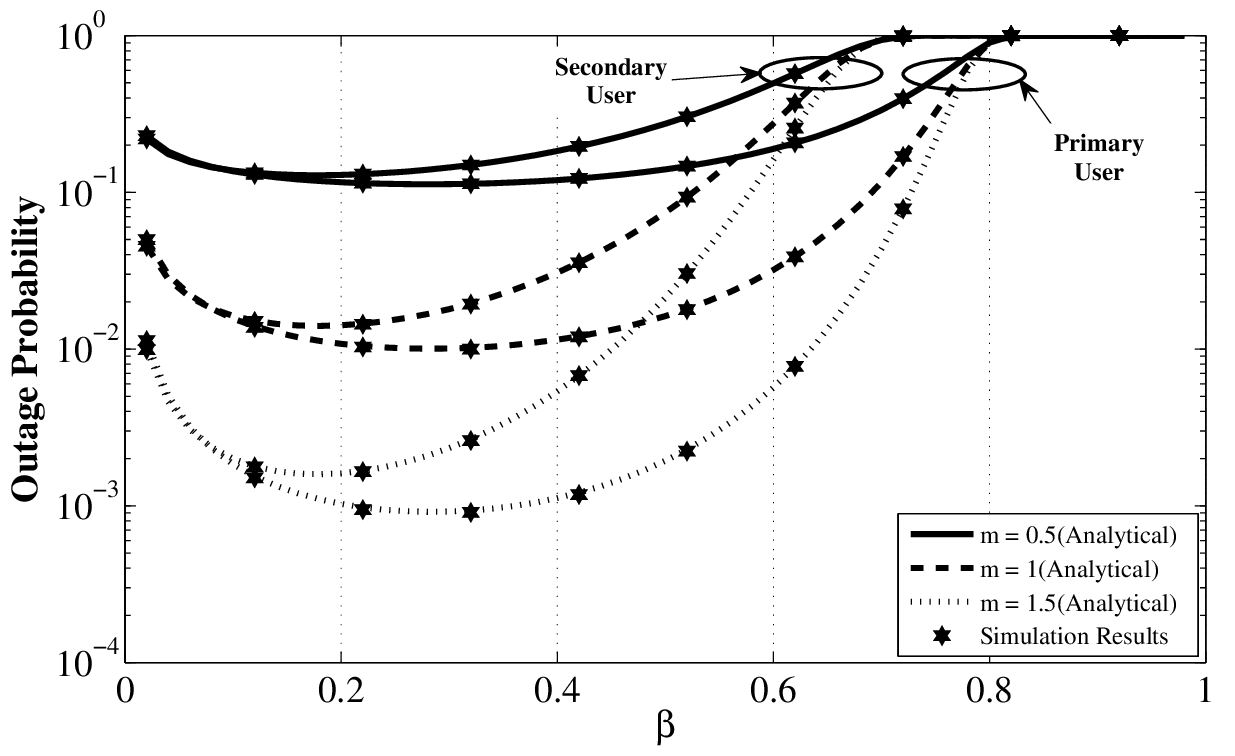}
        		\caption{}
        		\label{fig:vb1-af}
        	\end{subfigure}
        	\caption{Variation patterns of outage probability for primary and secondary users in TS-CSS protocol. The variation of outage in DF relaying w.r.t $\alpha$ have been shown in Fig.\ref{fig:va1-df}, and Fig.\ref{fig:vb1-df} shows the plot versus $\beta$. The respective variations in AF relaying have been plotted in Figs. \ref{fig:va1-af} and \ref{fig:vb1-af}.}
	  	\end{figure*}
	  	\begin{Prop}
	  	\label{thrm:2}
	  		The exact outage probabilities of the primary user, $\Pr_{out_1}^{\aThree}$, and that of the secondary user, $\Pr_{out_2}^{\aThree}$ in DF relaying mode for PS-CSS protocol can be derived in closed-form as 
	  			\begin{IEEEeqnarray*}{lll}
	  				\Pr_{out_1}^{\aThree} &=& 1-\left[(1-\Gamma(m,Y1))\right.\nonumber\\
	  						   && \left.\left(1-\frac{1}{\Gamma(m)^2}\,\MeijerG[\mu]{1}{1}{m,m}{0}{\frac{Z1}{\theta_1 \theta_2}}\right)\right], \IEEEeqnarraynumspace \IEEEyesnumber \IEEEyessubnumber \label{eq:pout2-df-1} \\
	  				\Pr_{out_2}^{\aThree} &=& 1-\left[(1-\Gamma(m,Y1))\right.\nonumber \\
	  						   && \left.\left(1-\frac{1}{\Gamma(m)^2}\,\MeijerG[\mu]{1}{1}{m,m}{0}{\frac{Z2}{\theta_1 \theta_3}}\right)\right], \IEEEeqnarraynumspace \IEEEyessubnumber \label{eq:pout2-df-2}
	  			\end{IEEEeqnarray*}%
	  			\textnormal{where},
	  			
	  			\begin{tabularx}{0.45\textwidth}{@{}XX@{}}
				  \begin{IEEEeqnarray}{lll}
				    \mathrm{Y1} &=& \tfrac{\psi_1 \sigma_{11}^2}{(1-\beta) \P_\p \theta_1}, \IEEEyesnumber \IEEEyessubnumber \label{eq:y1-ps-df}
	  				\\
	  				\mathrm{Z1} &=& \tfrac{\alpha \sigma_{22}^2 \psi_2}{\eta \P_\p \beta}, \IEEEyessubnumber \label{z1-ps-df}
	  				\\
	  				\mathrm{Z2} &=& \tfrac{(1-\alpha) \psi_3 \sigma_{32}^2}{\eta \P_\p\beta}, \IEEEyessubnumber \label{z2-ps-df}
				  \end{IEEEeqnarray} &
				  \begin{IEEEeqnarray}{lll}
	  				\psi_1 &=& 2^{{\frac{2 \R_\p}{\T}}}-1, \IEEEyessubnumber \label{eq:psi1-ps-df}
	  				\\
	  				\psi_2 &=& 2^{{\frac{2 \R_\p}{\alpha \T}}}-1, \IEEEyessubnumber \label{eq:psi2-ps-df}
	  				\\
	  				\psi_3 &=& 2^{{\frac{2 \R_\s}{(1-\alpha) \T}}}-1. \IEEEyessubnumber \label{eq:psi3-ps-df}
				  \end{IEEEeqnarray}
				\end{tabularx}
		\end{Prop}

	  	 \begin{IEEEproof}
	  	 Please, refer to Appendix~\ref{apndx3}.  
	  	 \end{IEEEproof}
	  	 
Note that, although at the first glance the forms of the outage expressions might look the same for both TS-CSS and PS-CSS scheme, that is not the case as the variables $\mathrm{Y1}$, $\mathrm{Z1}$ \text{and} $\mathrm{Z2}$ are different in each context.
	\subsection{AF Relaying Scheme}
	  In AF relaying mode, the signal received at ST in Phase 1, $y_{\ST}^{\,\,\prime}$, is given by \eqref{eq:yst-ps}. In Phase 2, the primary signal at ST is directly forwarded to PR for $\alpha \T/2$ time, and then ST transmits its own signal towards SR for the remaining slot time of $(1-\alpha)\T/2$. Therefore, the signal received at PR is given by
		\begin{equation}
			y_{\PR} = \frac{\sqrt{\P_{r_1^{\prime}}}}{\sqrt{(1-\beta)\P_\p \gamma_1 + \sigma_{11}^2}} y_{\ST}^{\,\,\prime} \h_2 + n_{22},
			\label{ypraf}
		\end{equation}
		where the factor $\sqrt{(1-\beta)\P_\p \gamma_1 + \sigma_{11}^2}$ in the denominator denotes the power constraint factor and $\P_{r_1}^{\prime}$ is given in \eqref{eq:prdf2-ps}. By substituting \eqref{eq:yst-ps} into \eqref{ypraf}, we get
		\begin{multline}
			y_{\PR} = \underbrace{\frac{\sqrt{\P_{r_1}^{\prime}\P_\p (1-\beta)}}{\sqrt{(1-\beta)\P_\p \gamma_1 + \sigma_{11}^2}}\h_1 \h_2}_\text{Signal Component} + \\ 
			 		  \underbrace{\frac{\sqrt{\P_{r_1}^{\prime}}}{\sqrt{(1-\beta)\P_\p\gamma_1+\sigma_{11}^2}}\h_2 n_{11}+n_{22}}_\text{Noise Component},
		\end{multline}
where, after the appropriate substitutions and arrangements, it follows that
		
		\begin{equation}
			\SNR_\PR^{\aFour}=\frac{\left(\displaystyle{\frac{\eta \beta \P_\p}{\alpha \sigma_{22}^2}}\gamma_1\gamma_2\right) \cdot
							 \left(\displaystyle{\frac{(1-\beta)\P_\p}{\sigma_{11}^2}}\gamma_1\right)}
							{\left(\displaystyle{\frac{\eta \beta \P_\p}{\alpha \sigma_{22}^2}}\gamma_1\gamma_2\right) + 
							 \left(\displaystyle{\frac{(1-\beta)\P_\p}{\sigma_{11}^2}}\gamma_1\right) + 1}.
							 \label{snrpr2-af}
		\end{equation}
		From \eqref{snrpr2-af}, the achievable rate at PR in AF mode can be written as 
		\begin{IEEEeqnarray}{rCl}
			\R_\PR^{\aFour} &=& \frac{\alpha \T}{2} \log_2(1+\SNR_\PR^{\aFour}).
			\label{eq:rpr-psaf}
		\end{IEEEeqnarray}
		The transmission towards SR in Phase 2 is independent of the mode of relaying used at ST. Thus, the signal received at SR, $y_{\SR}$, is given by \eqref{eq:ysr-ps} so that the instantaneous SNR at SR can be expressed as
		\begin{IEEEeqnarray}{rCl}
			\SNR_\SR^{\aFour} &=& \frac{\eta \beta \P_\p}{(1-\alpha) \sigma_{32}^2} \gamma_1 \gamma_3.
		\end{IEEEeqnarray}
		As in TS-CSS protocol, it always has transmission in Phase 2 in AF relaying scheme, hence making better utilization of the block time T. Thus, the achievable rate at SR is given by
		\begin{IEEEeqnarray}{rCl}
			\R_\SR^{\aFour} &=& \frac{(1-\alpha) \T}{2} \log_2(1+\SNR_\SR^{\aFour}).
		\end{IEEEeqnarray}
		\begin{Prop}
		\label{thrm-2-af}
			The exact outage probabilities of the primary user, $\Pr_{out_1}^{\aFour}$, and that of the secondary user, $\Pr_{out_2}^{\aFour}$ for AF relaying mode in PS-CSS protocol are given, respectively, by
			    \begin{IEEEeqnarray*}{lll}
	 				\Pr_{out_1}^{\aFour} &=&
	 				1-\int_\mathrm{Y1}^\infty \tfrac{\yy^{(m-1)} e^{-\yy}}{\Gamma(m)} \Gamma_u\left(m,\tfrac{\psi_1(\bb\theta_1 \yy + 1)}{\theta_1 \theta_2 (\aa\bb\theta_1 \yy^2 - \psi_1 \aa \yy)}\right) \mathrm{d}\yy, \\ \IEEEyesnumber \IEEEyessubnumber \IEEEeqnarraynumspace \label{eq:prout1-af-ps}\\
	 				\Pr_{out_2}^{\aFour} &=&
	 				\frac{1}{\Gamma(m)^2} \ \MeijerG[\mu]{1}{1}{m,m}{0}{\frac{Z2}{\theta_1 \theta_3}}, \IEEEeqnarraynumspace \IEEEyessubnumber \IEEEeqnarraynumspace \\
	 			\end{IEEEeqnarray*}%
	 		\textnormal{where},
	 		
	 		\begin{tabularx}{0.48\textwidth}{@{}XX@{}}
 					 \begin{IEEEeqnarray*}{lll}
  					  \aa &=& \tfrac{\eta \P_\p \beta}{\alpha \sigma_{22}^2}, \IEEEyesnumber \IEEEyessubnumber \label{a1-af}
  					  \\
  					  \mathrm{Y1} &=& \tfrac{\psi_1 \sigma{11}^2}{\P_\p \theta_1}, \IEEEyessubnumber \label{eq:y11-af}
	  				  \\ 
	 				  \mathrm{Z2} &=& \tfrac{(1-\alpha)\psi_2\sigma_{32}^2}{\eta \P_\p \beta}, \IEEEyessubnumber \label{eq:z21-af}
  					\end{IEEEeqnarray*}
  					&
  					\begin{IEEEeqnarray*}{lll}
  						\bb &=&\tfrac{(1-\beta)\P_\p}{\sigma_{11}^2}, \IEEEyessubnumber\label{b1-af}
 						 \\
	 					\psi_1 &=& 2^{{\frac{2 \R_\p}{\alpha \T}}}-1, \IEEEyessubnumber \label{eq:psi11-af}
	 					\\
	 					\psi_2 &=& 2^{{\frac{2 \R_\s}{(1-\alpha)\T}}}-1. \IEEEyessubnumber \label{eq:psi21-af}
  					\end{IEEEeqnarray*}
			\end{tabularx}
		\end{Prop}

		\begin{IEEEproof}
			Please, refer to Appendix~\ref{apndx4}.
		\end{IEEEproof}
		\begin{figure}[!tp]
	  	   		\centering
        		\includegraphics[width=0.5\textwidth]{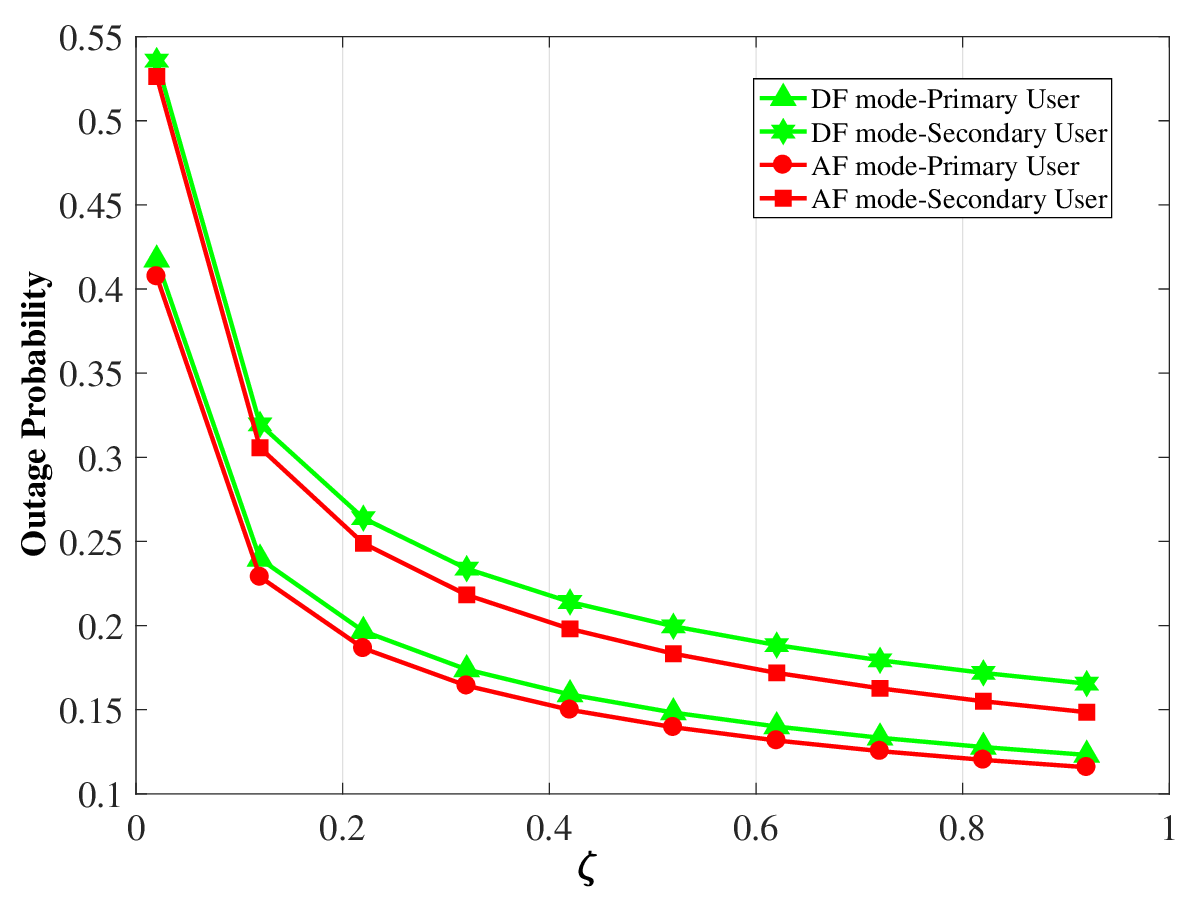}
        		\caption{Outage probability plot versus $\zeta$ for $\alpha=0.7 \: \text{and} \: \beta=0.3$ for both primary and secondary users in TS-CSS Protocol, $m$ = 1.}
        		\label{fig:ve1-df}
	  	   \end{figure}
	\section{Simulation results and Discussions} \label{sec:5}
In this Section, some representative numerical examples are depicted in order to study the effect of various system parameters on the outage performance. The analytical results have been validated through Monte Carlo simulations. The numerical simulations are carried by averaging over $10^6$ independent random realizations of the Nakagami-$m$ fading channels $\h_1, \h_2 \text{ and } \h_3$. The results for each protocol are plotted separately and various modes and protocols are compared with each other in terms of outage. Unless otherwise specified, the AWGN noise variance is taken to be equal along all channels, and the SNR, $\frac{\P_\p}{\sigma^2}=43$ dB. The path loss exponent $v$ = 3 which holds in cases concerning urban areas (\cite{meyrdigital,nasir2013relaying}). The nodes are placed such that the normalized distances between the links $\PT-\ST=\ST-\PR$ = 1 m and $\ST-\SR=\PT-\SR$ = 0.5 m, as in \cite{mousavifar2014wireless,vashistha2015outage}. The energy harvesting efficiency $\zeta$ is taken to be 1, and the threshold rates $\R_\p$ = $\R_\s$ = 1 bit/sec/Hz. The optimal values for $\alpha$ $(0<\alpha<1)$ and $\beta$ $(0<\beta<1)$ are those values, which minimize the system outage of primary and secondary systems. We note that the obtained results only show performance upper bounds, as many factors like packet loss, errors and practically achieved values of parameters which would further decrease the throughput are not considered here.
	  	\begin{figure}[!tp]
	  	   		\centering
        		\includegraphics[width=0.5\textwidth]{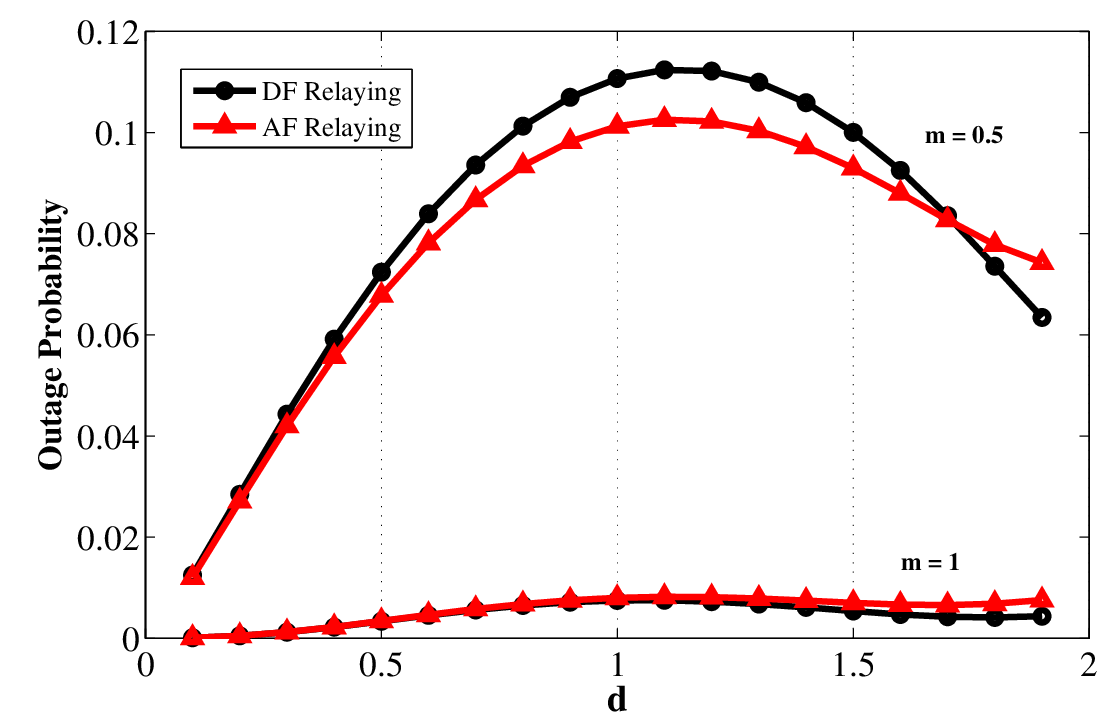}
        		\caption{Outage probability of the primary user versus $d$ in TS-CSS protocol ( $d_{PT-ST}=d$ and $d_{ST-PR}=2-d$).}
        		\label{fig:vd-af-df}
	  	\end{figure} 
		\begin{figure*}
	  	   		\centering
	  			\begin{subfigure}{0.5\textwidth}
	  				\centering
        			\includegraphics[width=\textwidth]{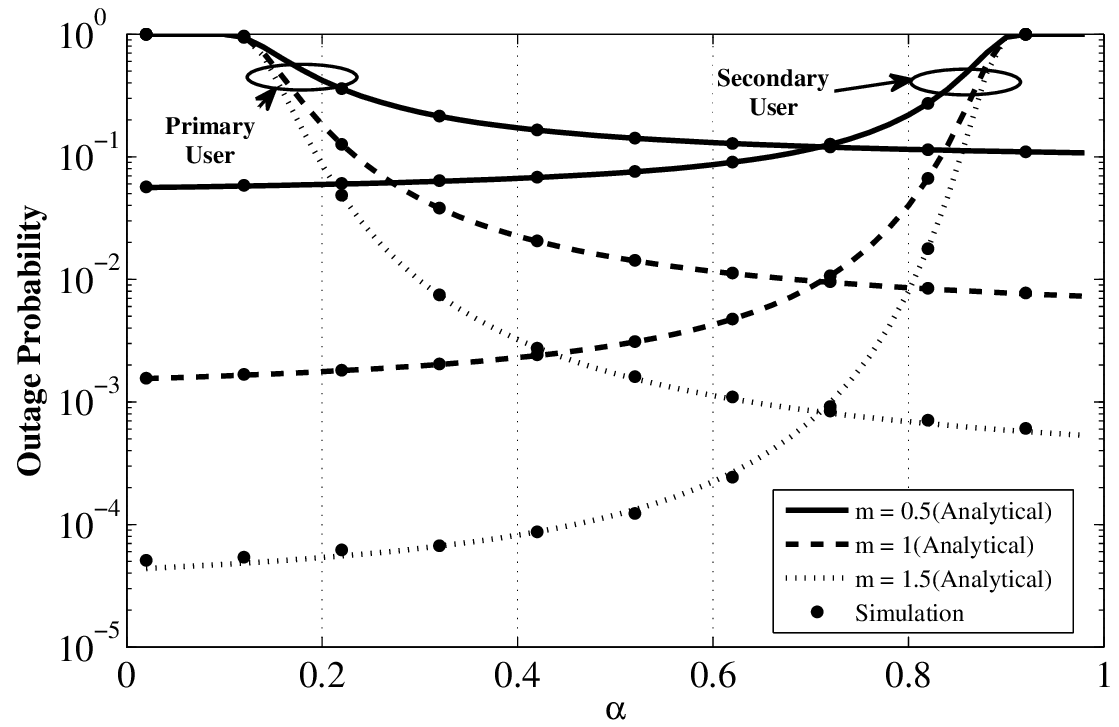}
        			\caption{}
        			\label{fig:va2-df}
	  	   		\end{subfigure}%
	  	   	~%
	  	   \begin{subfigure}{0.5\textwidth}
	  	   		\centering
        		\includegraphics[width=\textwidth]{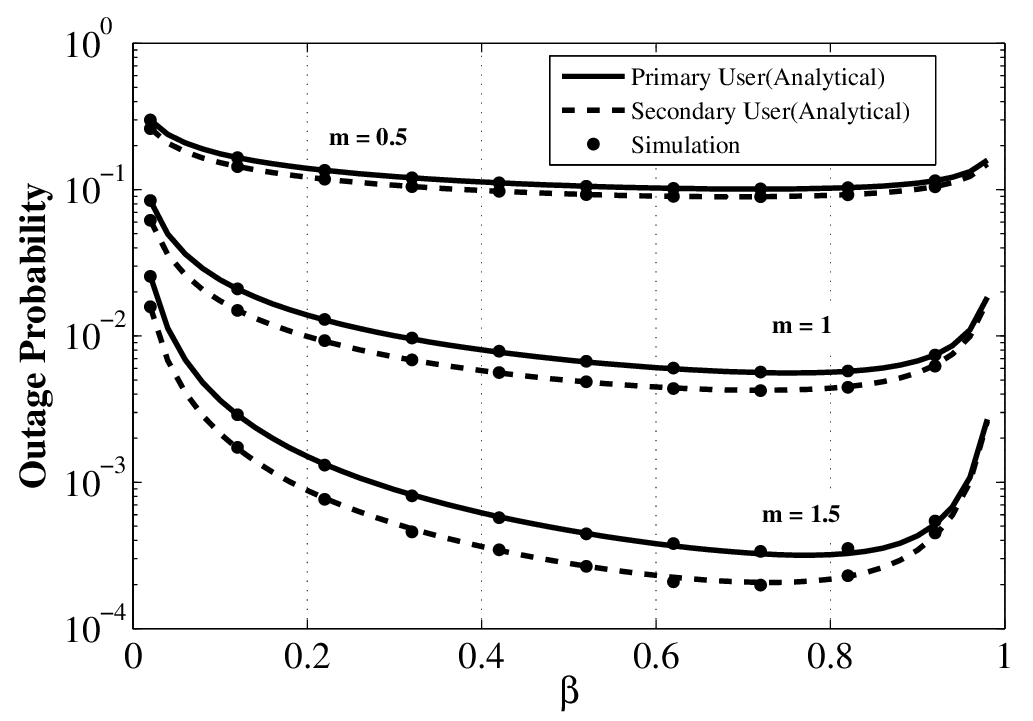}
        		\caption{}
        		\label{fig:vb2-df}
        	\end{subfigure}
        	\newline
        	\begin{subfigure}{0.5\textwidth}
	  				\centering
        			\includegraphics[width=\textwidth]{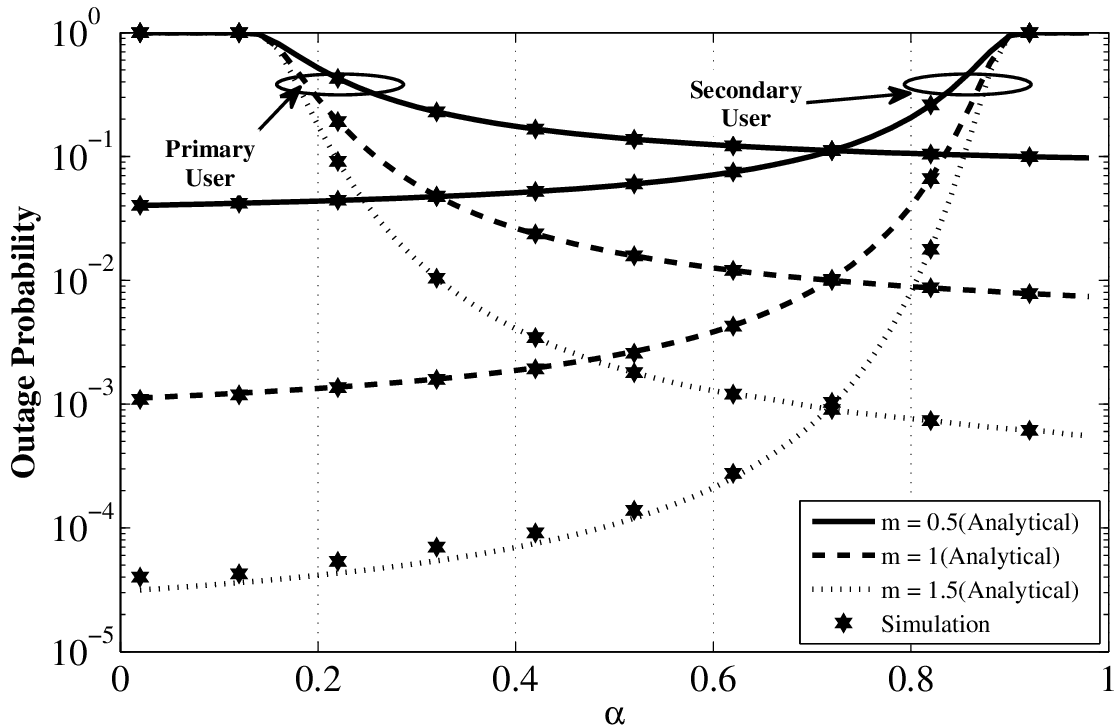}
        			\caption{}
        			\label{fig:va2-af}
	  	   		\end{subfigure}%
	  	   	~%
	  	   \begin{subfigure}{0.5\textwidth}
	  	   		\centering
        		\includegraphics[width=\textwidth]{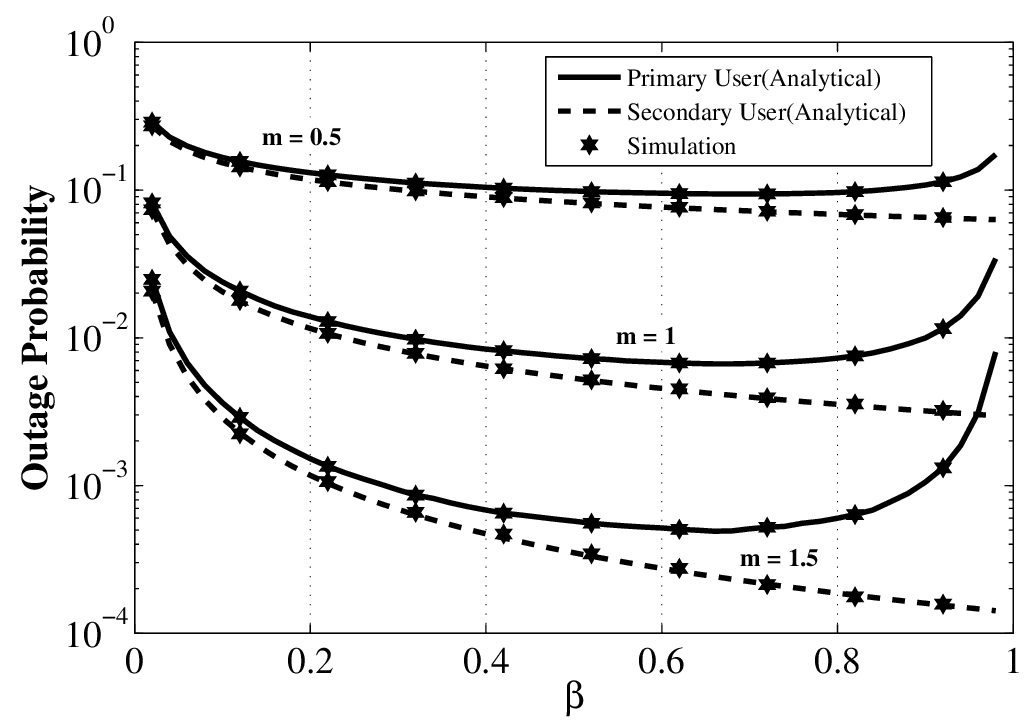}
        		\caption{}
        		\label{fig:vb2-af}
        	\end{subfigure}
        	\caption{Variation patterns of outage probability for primary and secondary users in PS-CSS protocol. Fig. \ref{fig:va2-df} shows the plot of outage versus $\alpha$ and Fig. \ref{fig:vb2-df} shows the plot versus $\beta$ in DF Relaying mode. The respective outage variations in AF Relaying mode are shown in Figs. \ref{fig:va2-af} and \ref{fig:vb2-af}.}
	  	\end{figure*}
		\subsection{TS-CSS Relaying Scheme}
		Fig. \ref{fig:va1-df} shows the variation of outage probability with respect to the time splitting ratio $\alpha$ for $\beta$ = 0.3 in DF relaying mode. Clearly the value of $\alpha$ at which the outages of primary and secondary are equal is independent of the value of $m$ (Nakagami fading parameter), and is equal to $0.65$. This value can be derived analytically by equating \eqref{eq:pout1-df} and \eqref{eq:pout2-df}. The variation patterns of outage with respect to $\beta$ are given in Fig. \ref{fig:vb1-df}, and the value of $\beta$ for which minimum outage is observed is around $0.3$ for $m$ = 1 and $m$ = 1.5 and around $0.25$ for $m$ = 0.5. 
		Fig. \ref{fig:vb1-df} also suggests that after exceeding a threshold value of $\beta$ around 0.75 for primary and 0.8 for secondary user, the outage goes to 1, or the throughput$\rightarrow$0. This can be explained as follows. For the primary user, it requires certain power threshold to transmit the signal, and once that much energy is harvested, increasing $\beta$ beyond this value would not give any additional advantage. Furthermore, when $\beta$ $> 0.8$, very less time is given for information decoding compared to energy harvesting, so the former becomes the dominating factor in deciding the outage. Also, larger $\beta$ implies lesser time for transmission at ST (refer Fig. \ref{fig:TS-CSS}). For secondary user, however, the time given for decoding at ST becomes the dominating factor rather earlier (at a lesser value of $\beta \approx 0.75$), because the distance between ST-SR is less than that of ST-PR, and thus it requires lesser energy threshold to transmit the signal to SR. So increasing $\beta$ beyond this value would ultimately prove detrimental. 

	  	Similar trend for AF relaying can be observed in Fig. \ref{fig:va1-af} against $\alpha$ for a value of $\beta$ = 0.3 and in Fig. \ref{fig:vb1-af} against $\beta$ for a value of $\alpha$ = 0.7. The shapes of the plots and threshold values suggest that the overall pattern in TS-CSS relaying scheme is similar irrespective of the mode of relaying used at ST for relaying data to PR. 

	  	The variation of the outage probability with the energy harvesting efficiency $\zeta$ for both the protocols for $m$ = 1 (Rayleigh fading) is shown in Fig. \ref{fig:ve1-df}. The graph demonstrates the fact that better energy harvesting circuits would improve the system performance to a great extent. Fig. \ref{fig:ve1-df} also establishes the marginal performance gain in AF relaying compared to DF relaying over a wide range of energy harvesting efficiency, $\zeta$.
		
	  	The distances between PT-ST and ST-PR are both taken to be 1m in each of the above results. However, an interesting plot in Fig. \ref{fig:vd-af-df} is obtained by varying the position of the relaying secondary transmitter and calculating the primary user outage in TS-CSS scheme in AF and DF relaying modes, for $\alpha$ = 0.7, $\beta$ = 0.3 and $\zeta$ = 1. The outage increases as ST is moved away from PT because as the distance between PT-ST increases, the received signal strength at ST will decrease owing to a larger path loss and hence larger outages at the destination. However, when the relaying ST is close to the destination PR, the outage again decreases as the energy harvested, though very less, is sufficient to transmit the signal to PR, located near ST. So, the ideal placement of the relay, even with a cooperative spectrum sharing mechanism, is close to PT, which agrees to the result in \cite{nasir2013relaying}. Another conclusion evident from the graph is that when ST is near to PT than PR, AF relaying gives higher throughput than DF relaying and vice-versa when ST is near PR, although the performance gain in either case is rather marginal. 
	\subsection{PS-CSS Relaying scheme}
		This protocol also showed similar behavior to that of the TS-CSS protocol in terms of the variation in the outage probability. Remarkably, comparison of Fig. \ref{fig:va1-df} and Fig. \ref{fig:va2-df} suggests that the variation of outage with respect to the time-splitting parameter $\alpha$ in both protocols follows the same pattern, for each value of $m$. This can be explained as follows. Both TS-CSS and PS-CSS protocol employ the parameter $\alpha$ as the time-sharing parameter in Phase 2, so they follow the same pattern of variation with respect to $\alpha$. Moreover, this pattern is also observed in AF mode, evident from Fig. \ref{fig:va2-df} and Fig. \ref{fig:va2-af}.

		On the other hand, the variation of $\beta$ in Fig. \ref{fig:vb2-df} in DF relaying does not quite follow the pattern as in Fig. \ref{fig:vb1-df}. The optimal value of $\beta$ is around 0.7 for all values of $m$. It is also clear that there is also no threshold $\beta$ as all values produce a non-zero throughput. The reason for this is as follows. The time splitting in transmission Phase 2 of the protocol is independent of the parameter $\beta$, so a larger $\beta$ would not alter the transmission time at ST (refer Fig. \ref{fig:PS-CSS}) and hence has lesser dominance over the outage at both secondary and primary user. The variation of energy harvesting efficiency $\zeta$ follows the same pattern in both the protocols, as more efficiency in both cases leads to lesser outage probability (Fig. \ref{fig:ve1-df}). 

		Another interesting observation is obtained by comparing Fig. \ref{fig:vb2-df} (DF) with Fig. \ref{fig:vb2-af} (AF). Although the variation is identically different from their TS-CSS counterparts, the outage probability of the secondary user in amplify forward mode is a monotonically decreasing function of $\beta$, suggesting that greater the value of $\beta$ chosen at ST, better will be the performance of the secondary user. This is because if $\beta$ is high, ST will get greater transmit power and hence better performance. This is compared with decode forward relaying, where increasing $\beta$ would in turn reduce the decoding capabilities at ST, and if ST has failed to decode the signal in Phase 1, it does not transmit its own signal too in phase 2, which is not the case with AF relaying.
		
\begin{figure}[!tp]
	  	   		\centering
        		\includegraphics[width=0.5\textwidth]{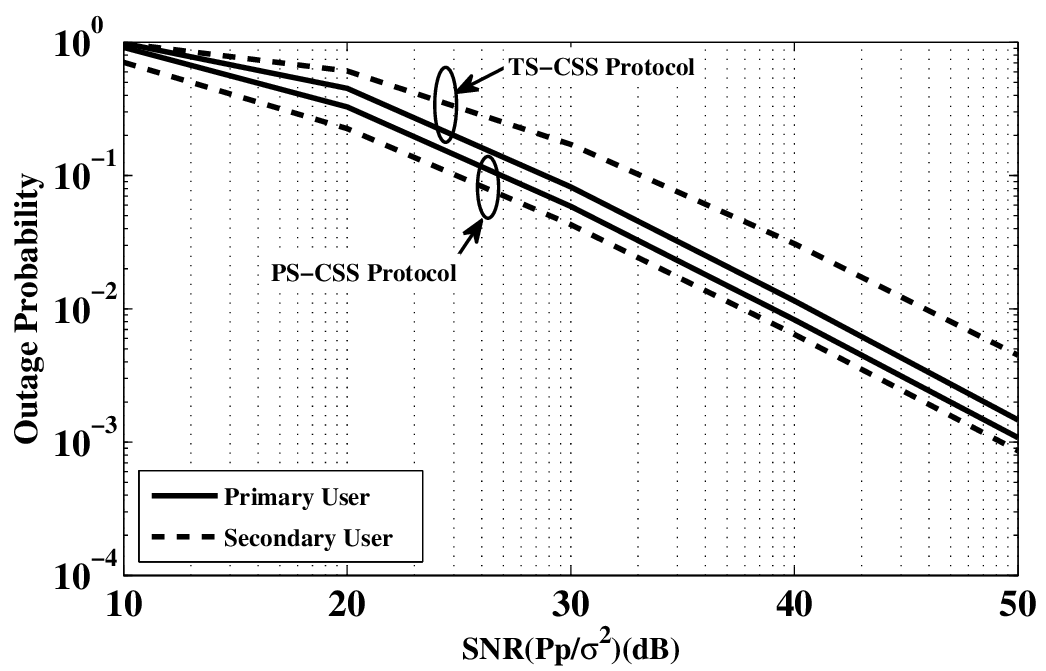}
        		\caption{Comparison of TS-CSS and PS-CSS relaying schemes for different SNR values in AF relaying. The values of $\alpha=0.7$, $\beta=0.6$, $\zeta=1$ and $m$ = 1.}
        		\label{fig:intercomp}
	  	\end{figure}

		Finally, both the relaying schemes are compared with each other over a wide range of SNRs and the plot is shown in Fig. \ref{fig:intercomp} for $m$ = 1 which corresponds to Rayleigh fading channel. Clearly, the PS-CSS relaying scheme outperforms the TS-CSS relaying scheme over a wide range of SNRs for both the primary and secondary users. So, we conclude that the power sharing based receiving has better throughput compared to time splitting receiving in a Rayleigh fading channel.

	\section{Conclusion}
	\label{sec:6}
		With an intention of addressing both the issues of energy constraint and spectrum efficiency in wireless networks, in this paper we proposed two protocols to carry out transmission in a spectrum sharing scenario between a primary user and an energy harvesting secondary user. Assuming an overlay mode, the outage probabilities of both protocols are derived, and it is found out that PS-based relaying outperforms TS-based relaying over a wide range of SNRs at optimum values of $\alpha$ and $\beta$. It is also found that AF relaying scheme is better than DF relaying for most cases except when ST is very near to PR, in which case the throughput of DF relaying has slightly better throughput. The explained protocols enable efficient usage of the spectrum without adversely affecting the performance of the primary user and promises longer lifetime of the sensor nodes through energy harvesting. 
	  	\bibliographystyle{IEEEtran} \bibliography{manuscript_27Nov}
	  	\numberwithin{equation}{section}
	\begin{appendices}
	\section{Proof for Proposition \ref{thrm:1}}
	\label{apndx1}
	\begin{IEEEproof}
	  	 	An outage is declared for the primary user if any of the links PT-ST or ST-PR fail to achieve the threshold rate required to decode the signal ($\R_\p$). For the secondary user, outage is dependent on the links PT-ST and ST-PR, and the decoding rate at SR must meet the threshold $\R_\s$. Mathematically speaking, we have that
	  	 		\begin{IEEEeqnarray*}{llll}
	  	 			& \Pr_{out_1}^{\aOne}&=& 1-[\Pr(\R_\ST^{\aOne} > \R_\p)\Pr(\R_\PR^{\aOne} > \R_\p)], \\ \IEEEyesnumber \IEEEyessubnumber \IEEEeqnarraynumspace \label{eq:pouts1-df} \\
	  	 		     &\Pr_{out_2}^{\aOne}&=& 1-[\Pr(\R_\ST^{\aOne} > \R_\p)\Pr(\R_\SR^{\aOne} > \R_\s)]. \\ \IEEEyessubnumber \IEEEeqnarraynumspace
	  	 			\label{eq:pouts2-df}
	  	 		\end{IEEEeqnarray*}\\
	  	    (i) Derivation of $\Pr(\R_\ST^{\aOne} > \R_\p$): Using \eqref{eq:snrst-df} for $\SNR_\ST^{\aOne}$, we get
        		\begin{IEEEeqnarray}{rRCl}
        			& \Pr(\R_\ST^{\aOne} > \R_\p) &=& \Pr\left(\SNR_\ST^{\aOne} > 2^{\tfrac{2\R_\p}{(1-\beta)\T}}-1\right),	\nonumber \\
         \Rightarrow& \Pr(\R_\ST^{\aOne} > \R_\p) &=& \Pr\left(\tfrac{\P_\p \gamma_1}{\sigma_{11}^2} > \psi_1 \right), \nonumber \\
         			&					 &=& \Pr\left(\gamma_1 > \tfrac{\psi_1 \sigma_{11}^2}{\P_\p}\right), \nonumber
        		\end{IEEEeqnarray}

	  	    where $\psi_1$ is given in \eqref{eq:psi1-df}. Then, knowing that
	  	    \begin{IEEEeqnarray}{rRCl}
	  	    	&\Pr(\gamma_1 > \mathrm{X}) &=& 1 - \Gamma\left(m,\frac{\mathrm{X}}{\theta_1}\right), \label{eq:gam_cdf}\\
	  	    	 & \Pr(\R_\ST^{\aOne}>\R_\p) &=& 1-\Gamma(m,\mathrm{Y1}), \nonumber
	  	    \end{IEEEeqnarray}
				
			where Y1 is given in \eqref{y1-df}.
			
	  	    (ii) Derivation of $\Pr(\R_\PR^{\aOne} > \R_\p)$: Using \eqref{eq:rpr} for $\R_\PR^{\aOne}$, we can write%
	  	     \begin{IEEEeqnarray*}{rrll}
	  	     	     & \Pr(\R_\PR^{\aOne} > \R_\p) &=& \Pr\left(\SNR_\PR^{\aOne} > 2^{\frac{2 \R_\p}{\alpha (1-\beta)\T}}-1\right), \nonumber \\
		  \Rightarrow& \Pr(\R_\PR^{\aOne} > \R_\p) &=& \Pr\left(\tfrac{2 \eta \P_\p \beta}{\alpha (1-\beta) \sigma_{22}^2}\gamma_1 \gamma_2>                                    \psi_2 \right),  \nonumber \\
					 &					  &=& \Pr(\gamma_1 \gamma_2 > \mathrm{Z1}), \nonumber   
	  	     \end{IEEEeqnarray*}
	  	     where $\psi_2$ is shown in \eqref{eq:psi2-df} and Z1 in \eqref{z1-df}. In \cite{karagiannidis2006bounds}, a closed-form expression for the CDF of the product of Gamma random variables was derived. Making use of those results, we have that
	  	     \begin{IEEEeqnarray}{lll}
	  	     	\Pr(\R_\PR^{\aOne} > \R_\p) &=& \Pr(\gamma_1 \gamma_2 > \mathrm{Z1}) \nonumber \\
	  	     					   &=& 1-\left(\frac{1}{\Gamma(m)^2}\,\MeijerG[\mu]{1}{1}{m,m}{0}{\frac{Z1}
	  	     					         {\theta_1\theta_2}}\right)\nonumber. \\
	  	     \end{IEEEeqnarray} 

	  	    (iii) Derivation of $\Pr(\R_\SR^{\aOne} > \R_\s)$: Using (9b) for  $\R_\SR^{\aOne}$, we can write
	  	    \begin{IEEEeqnarray}{rrll}
	  	     	     & \Pr(\R_\SR^{\aOne} > \R_\s) &=& \Pr\left(\SNR_\SR^{\aOne} > 2^{\frac{2 \R_\s}{(1-\alpha)(1-\beta)\T}}-1\right), \nonumber \\
		  \Rightarrow& \Pr(\R_\SR^{\aOne} > \R_\s) &=& \Pr\left(\tfrac{2 \eta \P_\p \beta}{(1-\alpha)(1-\beta) \sigma_{32}^2}\gamma_1 
		  									  \gamma_3 > \psi_3 \right), \nonumber \\
					&					  &=& \Pr(\gamma_1 \gamma_3 > \mathrm{Z2}), \nonumber    
	  	     \end{IEEEeqnarray}
	  	     where $\psi_3$ is given  in \eqref{eq:psi3-df} and Z2 in \eqref{z2-df}. Therefore,
	  	     \begin{IEEEeqnarray*}{lll}
	  	     	\Pr(\R_\SR^{\aOne} > \R_\s) &=& \Pr(\gamma_1 \gamma_3 > \mathrm{Z2}) \nonumber \\
	  	     					   &=& 1-\left(\frac{1}{\Gamma(m)^2}\,\MeijerG[\mu]{1}{1}{m,m}{0}{\frac{Z2}
	  	     					         {\theta_1 \theta_3}}\right)\nonumber. \\
	  	     \end{IEEEeqnarray*}
	  	      Finally, by substituting the above results in \eqref{eq:pouts1-df} and \eqref{eq:pouts2-df}, it results in \eqref{eq:pout1-df} and \eqref{eq:pout2-df} respectively. \qedhere
	  	      The value of $\mu_1$=1 in the above equations. 
	  	 \end{IEEEproof} 

	  	 \section{Proof for Proposition \ref{thm1-af}}
	  	 \label{apndx2}
	  	 \begin{IEEEproof}
	The outage probability in this case is defined as 
	\begin{IEEEeqnarray}{rCl}
		\Pr_{out_1}^{\aTwo} &=& \Pr(\R_\PR^{\aTwo} < \R_\p), \IEEEyesnumber \IEEEyessubnumber \label{eq:pout1-two}\\
		\Pr_{out_2}^{\aTwo} &=& \Pr(\R_\SR^{\aTwo} < \R_\s). \IEEEyessubnumber \label{eq:pout2-two}
	\end{IEEEeqnarray}
	From \eqref{eq:rpr-tsaf}, we can write $\Pr_{out_1}^{\aTwo} = \Pr(\R_\PR^{\aTwo}<\R_\p)$ as
	\begin{IEEEeqnarray*}{ll}
		& \Pr\left(\frac{\alpha(1-\beta)\T}{2}\log_2(1+\SNR_\PR^{\aTwo}) < \R_\p \right),  \IEEEnosubnumber\\
						  =& \Pr\left(\SNR_\PR^{\aTwo}<2^{\frac{2\R_\p}{\alpha(1-\beta)\T}}-1\right), \IEEEnosubnumber\\
						  =& \Pr\left(\frac{\left(\displaystyle{\frac{2 \eta \P_\p \beta}{\alpha(1-\beta)\sigma_{22}^2}\gamma_1 \gamma_2}\right)\left(\displaystyle{\frac{\P_\p}{\sigma_{11}^2}\gamma_1}\right)}{\left(\displaystyle{\frac{2 \eta \P_\p \beta}{\alpha(1-\beta)\sigma_{22}^2}\gamma_1 \gamma_2} \right) + \left(\displaystyle{\frac{\P_\p}{\sigma_{11}^2}\gamma_1}\right)+ 1} < \psi_1\right), \IEEEnosubnumber \IEEEeqnarraynumspace\\
						 =& \Pr\left(\frac{(\aa\gamma_1\gamma_2)(\bb\gamma_1)}{(\aa\gamma_1\gamma_2)+(\bb\gamma_1)+1}<\psi_1\right), \IEEEnosubnumber \IEEEeqnarraynumspace
	\end{IEEEeqnarray*}
	where a and b are given in \eqref{a-af} and \eqref{b-af} respectively. This results in
	\begin{IEEEeqnarray*}{lll}
		\Pr_{out_1}^{\aTwo} &=& \Pr(\frac{\aa\bb\gamma_1^2\gamma_2}{a\gamma_1\gamma_2+\bb\gamma_1+1}<\psi_1),  \IEEEnosubnumber\\
						  &=& \Pr(\aa\bb\gamma_1^2\gamma_2 < \aa\psi_1\gamma_1\gamma_2 + \bb\psi_1\gamma_1+\psi_1), \IEEEnosubnumber \IEEEeqnarraynumspace\\
						  &=& \Pr(\gamma_2(\aa\bb\gamma_1^2-\aa\psi_1\gamma_1)<\psi_1(\bb\gamma_1+1)), \IEEEnosubnumber \IEEEeqnarraynumspace
	\end{IEEEeqnarray*}
	which can be written as
	\begin{multline}
		\Pr_{out_1}^{\aTwo} = \Pr(\R_\PR^{\aTwo}<\R_\p)\\
			 =\int_0^\infty \textit{f}_{\gamma_1}(Z) \Pr(\gamma_2(\aa\bb Z^2 - \aa\psi_1 Z)<\psi_1(\bb Z+1)) \textrm{d}Z.
		\label{eq:int1-1}
	\end{multline}
	Consider now the term $\Pr(\gamma_2(\aa\bb Z^2 - \aa\psi_1 Z)<\psi_1(\bb Z+1))$ in (B.2). Let $\rho = \frac{\psi_1}{\bb}$. From \eqref{eq:int1-1}, if the factor $\rho >\gamma_1$, the left-hand side (LHS) of the expression is negative and right-hand side (RHS) is positive, resulting in $\Pr$(LHS $<$ RHS) = 1. Else, if $\rho < \gamma_1$, this quantity can be written as $\Pr(\gamma_2<\frac{\psi_1(\bb Z+1)}{(\aa\bb Z^2 - \aa\psi_1 Z)})$. Thus, \eqref{eq:int1-1} can be written as
	\begin{multline}
		\Pr_{out_1}^{\aTwo} = \int_0^\rho\, \left(\textit{f}_{\gamma_1}(Z) \Pr\left(\gamma_2>\frac{\psi_1(\bb Z+1)}{\aa\bb Z^2-\psi_1 \aa Z}\right)\textrm{d}Z\right) \\
					+ \int_\rho^\infty \left(\textit{f}_{\gamma_1}(Z)  \Pr\left(\gamma_2<\frac{\psi_1(\bb Z+1)}{\aa\bb Z^2-\psi_1 \aa Z}\right)\textrm{d}Z\right).
		\label{eq:int2-1}
	\end{multline}
	Knowing that $\gamma_1$ and $\gamma_2$ follow a Gamma distribution, (B.3) can be rewritten as
	\begin{align}
		\Pr_{out_1}^{\aTwo} = 1 - \int_\rho^\infty\, \frac{Z^{m-1}e^{-Z/\theta_1}}{\Gamma(m)\theta_1^m} \Gamma_u\left(\tfrac{\psi_1(\bb Z+1)}{\theta_2(\aa\bb Z^2 - \psi_1 \aa Z)}\right) \textrm{d}Z. \nonumber  \\
		\label{outZ}
	\end{align}

	By substituting $Z/\theta_1=y$ in \eqref{outZ} gives us the desired form for $\Pr_{out_1}^{\aTwo}$ as presented in \eqref{eq:pout1-af}. Due to the presence of higher order terms in the upper normalized incomplete Gamma function in \eqref{outZ}, the integration cannot be further simplified and hence is solved numerically in the simulation results. 

	The outage of the secondary user is given in \eqref{eq:pout2-two} and is similar to the calculation in proof for Proposition \ref{thrm:1}. The value of $\mu_1 = 1$ in the above equations. \qedhere

\end{IEEEproof}

	  	 \section{Proof for Proposition \ref{thrm:2}}
	  	 \label{apndx3}
	  	   	 \begin{IEEEproof}
	  	 	In the PS-CSS protocol too, the outage probabilities are defined as
	  	 	\begin{IEEEeqnarray*}{llll}
	  	 			& \Pr_{out_1}^{\aTwo}&=& 1-[\Pr(\R_\ST^{\aTwo} > \R_\p)\Pr(\R_\PR^{\aTwo} > \R_\p)], \\ \IEEEyesnumber \IEEEyessubnumber \IEEEeqnarraynumspace \label{eq:ps-df-pout1} \\
	  	 		     &\Pr_{out_2}^{\aTwo}&=& 1-[\Pr(\R_\ST^{\aTwo} > \R_\p)\Pr(\R_\SR^{\aTwo} > \R_\s)].\\ \IEEEyessubnumber \IEEEeqnarraynumspace
	  	 		     \label{eq:ps-df-pout2}
	  	 		\end{IEEEeqnarray*}%

	  	 	Next, we evaluate the expression term by term.
	  	 	
	  	    (i) Derivation of $\Pr(\R_\ST^{\aThree} > \R_\p$): Using $\SNR_\ST^{\aThree}$ from \eqref{eq:snr-st-2}, we get
        		\begin{IEEEeqnarray}{rrll}
        			& \Pr(\R_\ST^{\aThree} > \R_\p) &=& \Pr\left(\tfrac{\T}{2}\log_2(1+\SNR_\ST^{\aThree}) > \R_\p \right), \nonumber \\
        			&                    &=& \Pr\left(\SNR_\ST^{\aThree} > 2^{\tfrac{2\R_\p}{\T}}-1\right),	\nonumber \\
         \Rightarrow& \Pr(\R_\ST^{\aThree} > \R_\p) &=& \Pr\left(\tfrac{(1-\beta)\P_\p \gamma_1}{\sigma_{11}^2} > \psi_1 \right), \nonumber \\
         			&					 &=& \Pr\left(\gamma_1 > \tfrac{\psi_1 \sigma_{11}^2}{\P_\p(1-\beta)}\right), \nonumber
        		\end{IEEEeqnarray}

	  	    where $\psi_1$ is given in \eqref{eq:psi1-ps-df}, yielding
	  	    \begin{IEEEeqnarray}{rlll}
	  	    	 & \Pr(\R_\ST^{\aThree}>\R_\p) &=& 1-\Gamma(m,\mathrm{Y1}),
	  	    \end{IEEEeqnarray}
				
			where Y1 is given in \eqref{eq:y1-ps-df}.
			
	  	    (ii) Derivation of $\Pr(\R_\PR^{\aThree} > \R_\p)$: Using \eqref{eq:raf} for $\R_\PR^{\aThree}$, we can write
	  	     \begin{IEEEeqnarray}{rrll}
	  	     	     & \Pr(\R_\PR^{\aThree} > \R_\p) &=& \Pr\left(\tfrac{\alpha \T}{2}\log_2(1+\SNR_\PR^{\aThree}) > \R_\p \right), \nonumber \\
	  	     	     &                    &=& \Pr\left(\SNR_\PR^{\aThree} > 2^{\frac{2 \R_\p}{\alpha \T}}-1\right), \nonumber \\
		  \Rightarrow& \Pr(\R_\PR^{\aThree} > \R_\p) &=& \Pr\left(\tfrac{\eta \P_\p \beta}{\alpha \sigma_{22}^2}\gamma_1 \gamma_2>                                    \psi_2 \right),  \nonumber \IEEEeqnarraynumspace \\
					 &					  &=& \Pr(\gamma_1 \gamma_2 > \mathrm{Z1}), \nonumber    
	  	     \end{IEEEeqnarray}
	  	     where $\psi_2$ is given in \eqref{eq:psi2-ps-df} and Z1 in \eqref{z1-ps-df}, yielding
	  	     \begin{IEEEeqnarray}{lll}
	  	     	\Pr(\R_\PR^{\aThree} > \R_\p) &=& \Pr(\gamma_1 \gamma_2 > \mathrm{Z1}), \nonumber \\
	  	     					   			   &=& 1-\left(\frac{1}{\Gamma(m)^2}\,\MeijerG[\mu]{1}{1}{m,m}{0}{\frac{Z1}
	  	     					                   {\theta_1\theta_2}}\right),\nonumber \\
	  	     \end{IEEEeqnarray} where $\mu_1$=1. 

	  	    (iii) Derivation of $\Pr(\R_\SR^{\aThree} > \R_\s$): Using \eqref{eq:raf} for $\R_\SR^{\aThree}$, we get
	  	    \begin{IEEEeqnarray}{rrll}
	  	     	     & \Pr(\R_\SR^{\aThree} > \R_\s) &=& \Pr\left(\tfrac{(1-\alpha)\T}{2}\log_2(1+\SNR_\SR^{\aThree}) > \R_\s\right), \nonumber \\
	  	     	     &                    &=& \Pr\left(\SNR_\SR^{\aThree} > 2^{\frac{2 \R_\s}{(1-\alpha)\T}}-1\right), \nonumber \\
		  \Rightarrow& \Pr(\R_\SR^{\aThree} > \R_\s) &=& \Pr\left(\tfrac{\eta \P_\p \beta}{(1-\alpha)\sigma_{32}^2}\gamma_1 
		  									  \gamma_3 > \psi_3 \right),  \nonumber \IEEEeqnarraynumspace \\
					&					  &=& \Pr(\gamma_1 \gamma_3 > \mathrm{Z2}), \nonumber    
	  	     \end{IEEEeqnarray}
	  	     where $\psi_3$ is given in \eqref{eq:psi3-ps-df} and Z2 in \eqref{z2-ps-df}. Therefore,
	  	     \begin{IEEEeqnarray}{lll}
	  	     	\Pr(\R_\SR^{\aThree} > \R_\s) &=& \Pr(\gamma_1 \gamma_3 > \mathrm{Z2}), \nonumber \\
	  	     					   &=& 1-\left(\frac{1}{\Gamma(m)^2}\,\MeijerG[\mu]{1}{1}{m,m}{0}{\frac{Z2}
	  	     					         {\theta_1 \theta_3}}\right),\nonumber \\
	  	     \end{IEEEeqnarray}
	  	     where $\mu_1$=1. Substituting the above results in \eqref{eq:ps-df-pout1} and \eqref{eq:ps-df-pout2} gives the relations for $\Pr_{out_1}^{\aThree}$ and $\Pr_{out_2}^{\aThree}$ as in \eqref{eq:pout2-df-1} and \eqref{eq:pout2-df-2} respectively. \qedhere
	  	 \end{IEEEproof}
	  	 \section{Proof for Proposition \ref{thrm-2-af}}
	  	 \label{apndx4}
	  	 \begin{IEEEproof}
	The outage probability in this case is defined in the same way as in AF relaying in TS-CSS protocol. 
	\begin{IEEEeqnarray}{rCl}
		\Pr_{out_1}^{\aFour} &=& \Pr(\R_\PR^{\aFour} < \R_\p), \IEEEyesnumber \IEEEyessubnumber \label{eq:pout1-four}\\
		\Pr_{out_2}^{\aFour} &=& \Pr(\R_\SR^{\aFour} < \R_\s). \IEEEyessubnumber \label{eq:pout2-four}
	\end{IEEEeqnarray}
	From \eqref{eq:rpr-psaf}, we can write $\Pr_{out_1}^{\aFour} = \Pr(\R_\PR^{\aFour}<\R_\p)$ as
	\begin{IEEEeqnarray*}{ll}
		& \Pr(\frac{\alpha \T}{2}\log_2(1+\SNR_\PR^{\aFour})<\R_\p),  \IEEEnosubnumber\\
						  =& \Pr(\SNR_\PR^{\aFour}<2^{\frac{2\R_\p}{\alpha \T}}-1), \IEEEnosubnumber\\
						  =& \Pr\left(\frac{\left(\displaystyle{\tfrac{\eta \beta \P_\p}{\alpha \sigma_{22}^2}}\gamma_1\gamma_2\right)
							 \left(\displaystyle{\tfrac{(1-\beta)\P_\p}{\sigma_{11}^2}}\gamma_1\right)}
							{\left(\displaystyle{\tfrac{\eta \beta \P_\p}{\alpha \sigma_{22}^2}}\gamma_1\gamma_2\right) + 
							 \left(\displaystyle{\tfrac{(1-\beta)\P_\p}{\sigma_{11}^2}}\gamma_1\right) + 1} < \psi_1\right), \IEEEnosubnumber \IEEEeqnarraynumspace\\
						 =& \Pr(\frac{(\aa\gamma_1\gamma_2)(\bb\gamma_1)}{(\aa\gamma_1\gamma_2)+(\bb\gamma_1)+1}<\psi_1), \IEEEnosubnumber \IEEEeqnarraynumspace
	\end{IEEEeqnarray*}
	where a and b are given in \eqref{a1-af} and \eqref{b1-af} respectively. The expression is similar to that of AF relaying in TS-CSS protocol. Following the same procedure, the outage probability can be given as
	
	\begin{align}
		\Pr_{out_1}^{\aFour} = 1 - \int_\rho^\infty\, \frac{Z^{m-1}e^{-Z/\theta_1}}{\Gamma(m)\theta_1^m} \Gamma_u\left(\tfrac{\psi_1(\bb Z+1)}{\theta_2(\aa\bb Z^2 - \psi_1 \aa Z)}\right) \textrm{d}Z. \nonumber \\
	\end{align}

	The substitution $Z/\theta_1=y$ gives us the desired result for $\Pr_{out_1}^{\aFour}$ as an integration in \eqref{eq:prout1-af-ps}. The integration cannot be further simplified and is solved numerically in the simulation results. 

	The outage of the secondary user is given in \eqref{eq:pout2-four} and is similar to the calculation in proof for Proposition~\ref{thrm:2}.\qedhere
\end{IEEEproof}

	\end{appendices}
\end{document}